\documentclass{aa}         
\usepackage{txfonts}
\usepackage{graphicx}
\bibpunct{(}{)}{;}{a}{}{,}      

\begin{document}

\title{Heating by transverse waves in simulated coronal loops}

\author{K. Karampelas\inst{1} \and T. Van Doorsselaere\inst{1} \and P. Antolin\inst{2}} 

\institute{Centre for mathematical Plasma Astrophysics, Department of Mathematics, KU Leuven, Celestijnenlaan 200B, 3001 Leuven, Belgium \linebreak \email{kostas.karampelas@kuleuven.be} \and School of Mathematics and Statistics, University of St. Andrews, St. Andrews, Fife KY16 9SS, UK} 

\date{Received <date> /
Accepted <date>}

\abstract{Recent numerical studies of oscillating flux tubes have established the significance of resonant absorption in the damping of propagating transverse oscillations in coronal loops. The nonlinear nature of the mechanism has been examined alongside the Kelvin-Helmholtz instability, which is expected to manifest in the resonant layers at the edges of the flux tubes. While these two processes have been hypothesized to heat coronal loops through the dissipation of wave energy into smaller scales, the occurring mixing with the hotter surroundings can potentially hide this effect.}
{We aim to study the effects of wave heating from driven and standing kink waves in a coronal loop.}
{Using the MPI-AMRVAC code, we perform ideal, three dimensional magnetohydrodynamic (MHD) simulations of both (a) footpoint driven and (b) free standing oscillations in a straight coronal flux tube, in the presence of numerical resistivity.}
{We have observed the development of Kelvin-Helmholtz eddies at the loop boundary layer of all three models considered here, as well as an increase of the volume averaged temperature inside the loop. The main heating mechanism in our setups was Ohmic dissipation, as indicated by the higher values for the temperatures and current densities located near the footpoints. The introduction of a temperature gradient between the inner tube and the surrounding plasma, suggests that the mixing of the two regions, in the case of hotter environment, greatly increases the temperature of the tube at the site of the strongest turbulence, beyond the contribution of the aforementioned wave heating mechanism.}
{}


\maketitle                      

\titlerunning{Heating by propagating transverse waves in simulated coronal loops}
\authorrunning{K. Karampelas et al.}

\section{introduction}

Since the discovery of transverse magnetohydrodynamic (MHD) oscillations \citep{aschwanden1999, nakariakov1999}, they have been the topic of many studies, both observational and numerical. Stretching from the lower chromosphere up to the solar corona, the physical characteristics of loops allow them to act as waveguides, effectively transferring energy between those different layers. Analytical studies on the nature of the transverse oscillations in cylindrical flux tubes \citep{zajtsev1975, ryutov1976, edwin1983wave} have described the different modes expected in a non-uniform plasma with cylindrical symmetry. 

Observations from the Coronal Multi-channel Polarimeter (CoMP), the Solar Dynamics Observatory (SDO) spacecraft and Hinode Solar Observatory have revealed the existence of ubiquitous transverse perturbations traveling along coronal loops, prominence threads and greater areas of the corona \citep{tomczyk2007, okamoto2007, tomczyk2009, mcintosh2011, nistico2013, anfinogentov2015}. Due to their high speeds and apparent incompressible nature, \citet{tomczyk2007} have considered these perturbations to be Alfv\'{e}n waves, traveling in the solar corona. Considering the energy budget of these propagating waves, \citet{tomczyk2007} estimated an energy flux four orders of magnitude smaller than needed to balance the radiative losses of the quiet solar corona. However, there has been a lot of uncertainty regarding the estimated energy carried by the waves in the solar atmosphere \citep{depontieu2007,mcintosh2011,goossens2013energyApJ,tvd2014energyApJ,thurgood2014ApJ,morton2016ApJ},
 with the line of sight (LOS) being a particularly important factor \citep{mcintosh2012,demoortel2012ApJ}. Meanwhile, the nature of these oscillations has been under debate, with \citet{tvd2008detection} proposing that they are in fact Alfv\'{e}nic, transverse surface (kink) waves, since they have been observed traveling along flux tubes in the solar atmosphere, rather than in a homogeneous plasma as it would be expected of Alfv\'{e}n waves. The  Alfv\'{e}nic nature of those kink waves in magnetic flux tubes has also been proven in \citet{goossens2009kinkftube}.
 
For the proposed heating of the solar atmosphere a dissipation mechanism is necessary for the observed oscillations to transfer their kinetic energy into internal energy of the plasma. \citet{tomczyk2009} reported significant spatial attenuation in the power of the aforementioned observed propagating waves. Additional observational evidence was presented in \citet{verthterradas2010}, while in \citet{terradas2010}, the mechanism of resonant absorption was used to analytically explain this spatial attenuation. The analogous mechanisms of resonant absorption \citep{sakurai1991, goossens1992linear, goossens1992resonant, goossens2002coronal, ruderman2002damping, arregui2005resonantly, goossens2006damping, goossens2011resonant} and mode coupling \citep{pascoe2010, pascoe2012,  demoortel2016} have been thus considered responsible for the damping of transverse waves in flux tubes. Through resonance, the energy of the global kink mode is transferred to local azimuthal Alfv\'{e}n modes in the boundary layer at the loop edges, reducing the amplitude of the oscillations. In the case in which multiple frequencies are excited (for example, should a broad-band driver be considered or in the case of non-linear effects due to resistivity and viscosity), smaller scales are created through the mechanism of phase mixing \citep{heyvaerts1983,poedts1996,soler2015}. The connection between resonant absorption and the heating of loops has been studied in the past \citep{ofman1994heat,ofman1994bheatvis,poedts1996,ofman1998}, where resistivity and/or viscosity were considered, in order to dissipate the energy contained into the created smaller scales. 

While studying the damping mechanisms for Alfv\'{e}n waves in the boundary layers of flux tubes, \citet{heyvaerts1983} predicted the existence of Kelvin-Helmholtz Instabilities (KHI) in the resonant layer, through a nonlinear connection with phase mixing. They argued that the strong shear velocities generated by the azimuthal Alfv\'{e}n waves can give rise to Kelvin-Helmholtz turbulence, which in turn reinforces the effects of phase mixing through the creation of smaller scales. Propagating waves were predicted to be Kelvin-Helmholtz stable, while standing oscillations should be unstable near the positions of the velocity antinodes. \citet{zaqarashvili2015ApJ} have also predicted that the higher values of azimuthal velocities at loop edges near these velocity antinodes would make standing kink modes and torsional Alfv\'{e}n waves Kelvin-Helmholtz unstable. The presence of KHI and its connection to turbulence has also been studied in chromospheric jets \citep{kuridze2015,kuridze2016} as an explanation of the observed, non-thermal, line broadening.

Three dimensional simulations in straight flux tubes confirmed the non-linear connection between resonant absorption, phase mixing and  KHI for driver generated azimuthal Alfv\'{e}n waves \citep{uchimoto1991,ofman1994nonlinear,poedts1997a,poedts1997b}. More recent numerical studies \citep{terradas2008,antolin2014fine,antolin2015resonant,antolin2016,magyar2015,magyar2016damping} have confirmed the development of Kelvin-Helmholtz induced turbulence in straight flux tubes even  for small amplitude standing kink waves. In particular, \citet{antolin2014fine}, through the use of forward modelling, proved that KHI can create apparent strands along flux tubes, as a LOS effect, providing us with a potential method to indirectly observe this instability in coronal loops. Additionally, \citet{magyar2016damping} showed that the developed KHI can lead to faster damping of standing transverse waves than analytically predicted from resonant absorption, further proving its effectiveness. 

Following the idea that resistive and viscous dissipation contributed to heating, \citet{antolin2014fine} suggested that the developed Kelvin-Helmholtz instabilities at the resonant layer could lead to an increase in the flux tube temperature. The profiles of his tube cross-section at the antinode position revealed the existence of small length scales in $z$-current density, similar to those created by the Kelvin-Helmholtz eddies. However, \citet{magyar2016damping} pointed out that the mixing between the colder flux tube and the hotter surrounding plasma led to higher average temperatures than those expected from the increase of the internal energy for the given simulation time.

In the current work, we focus on the temperature evolution in both driven and freely transversely  oscillating flux tubes,  due to nonlinear
dissipation of wave energy. Physical resistivity is not included in our three dimensional ideal MHD models, but the effects of numerical resistivity are present and are used in the study of wave energy dissipation. We consider models of equal temperature inside and outside of the flux tube, so that we can isolate the mechanisms of wave heating from the effects of mixing between regions of different temperature. The effects of mixing are considered in a third case, where we introduced a temperature gradient across the tube axis. We briefly discuss the dynamics of our driven oscillating tubes, as well as the implication of their dynamical evolution on the spatial evolution of the loop heating. 

\section{Numerical models}
\subsection{Equilibrium}

\begin{figure}
\centering
\includegraphics[trim={1cm 3cm 10.5cm 0.9cm},clip,scale=0.16]{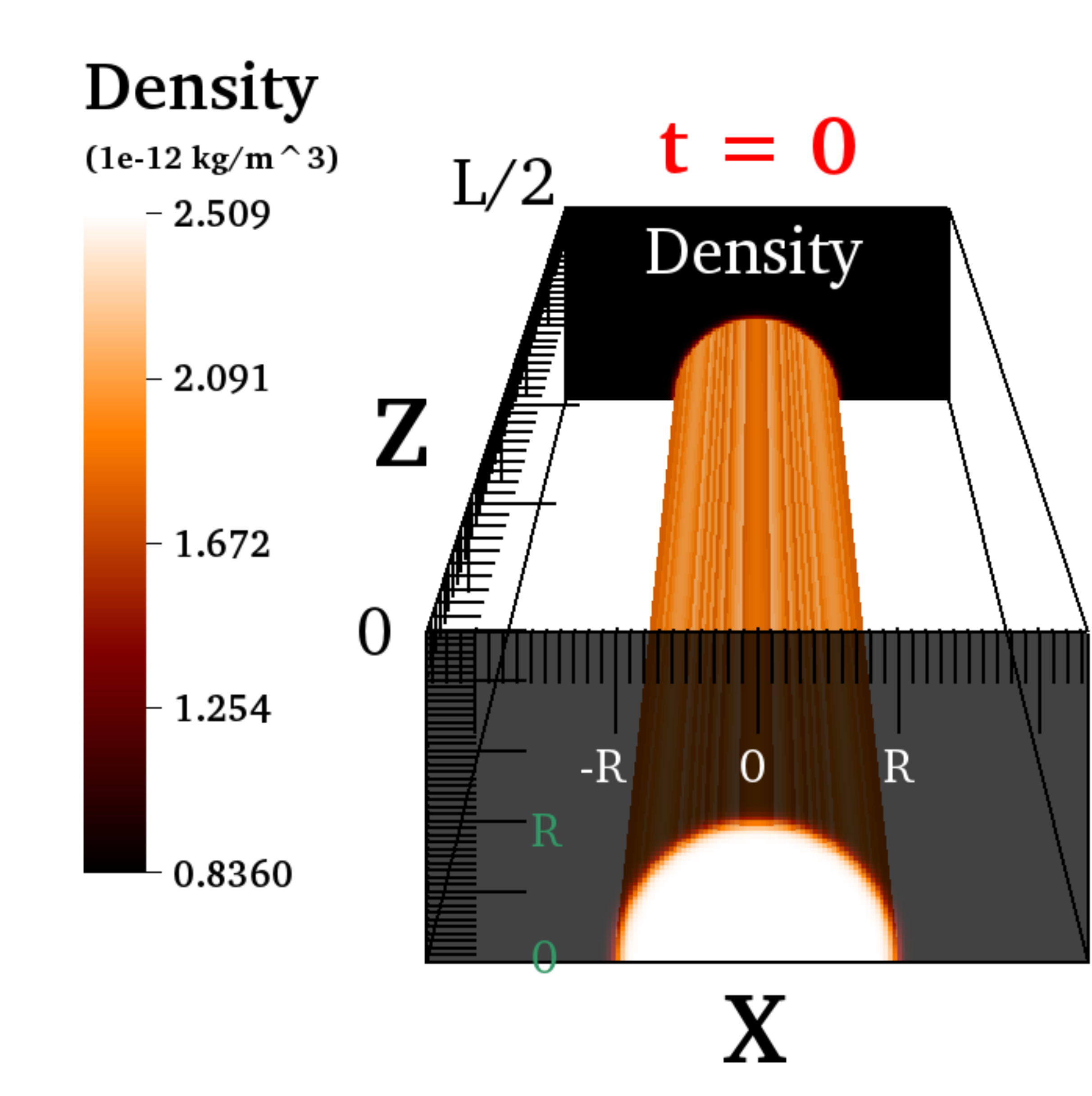}
\includegraphics[trim={9cm 3cm 8cm 0.9cm},clip,scale=0.16]{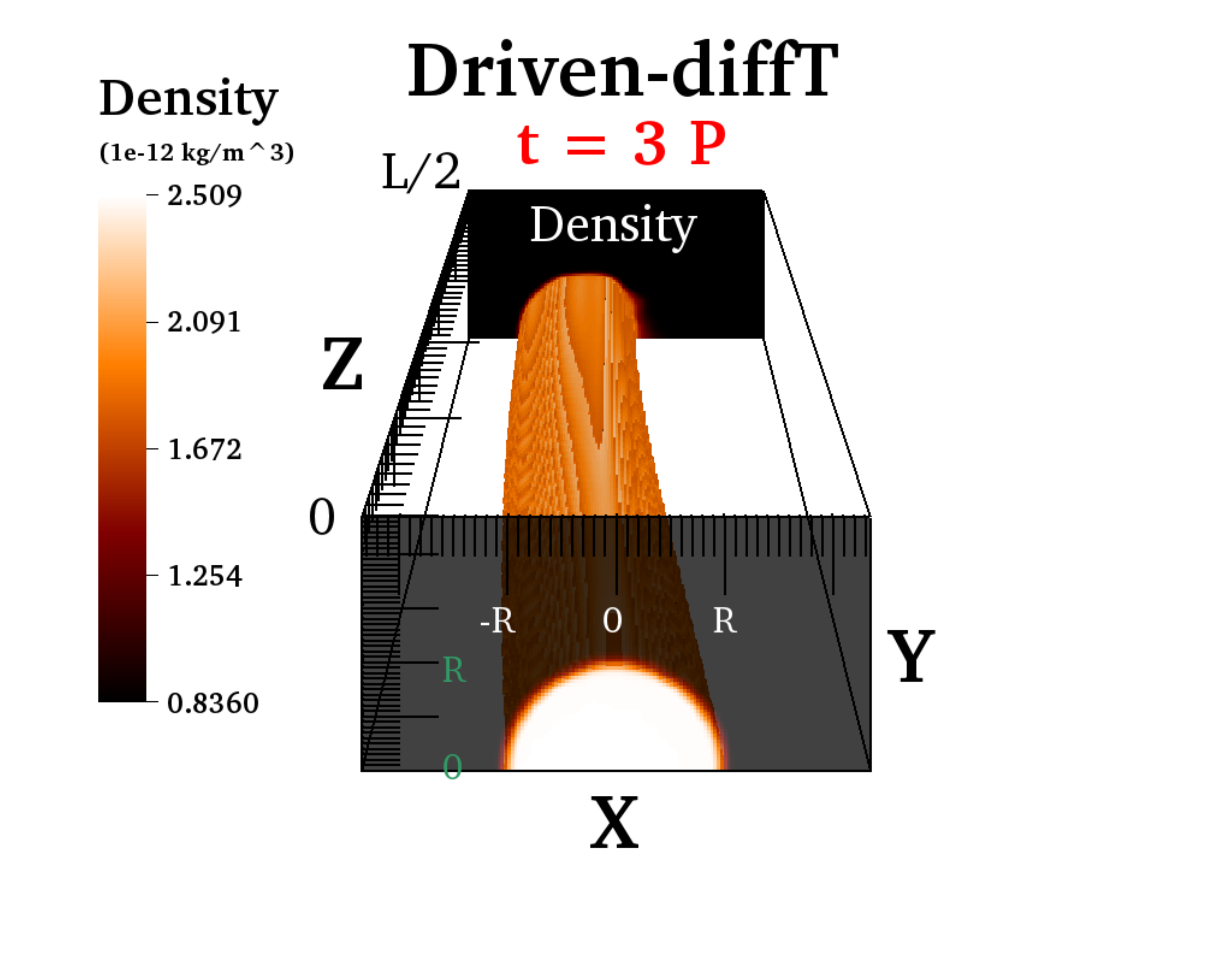}
\includegraphics[trim={1cm 3cm 10.5cm 0.9cm},clip,scale=0.16]{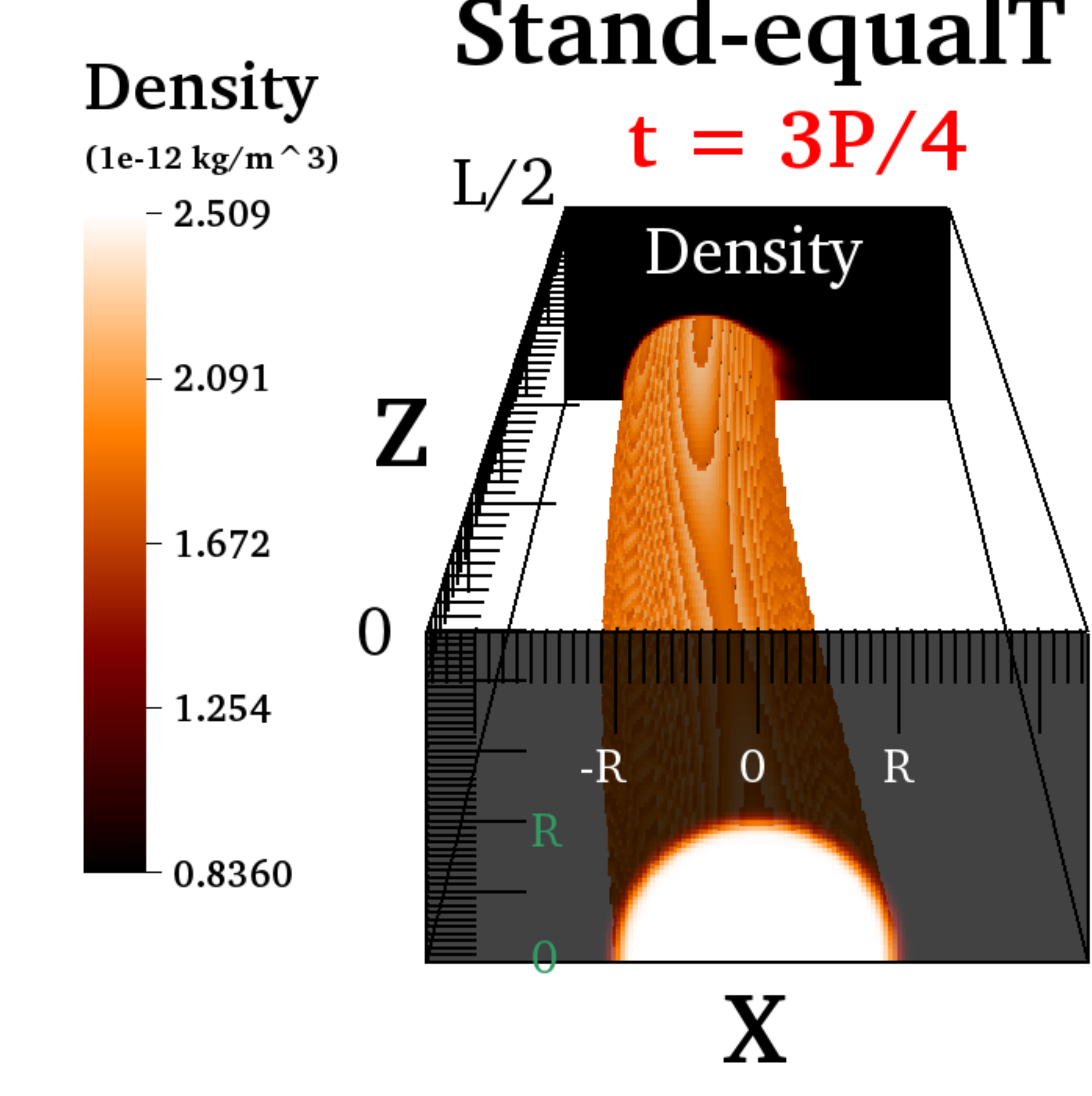}
\includegraphics[trim={9cm 3cm 8cm 0.9cm},clip,scale=0.16]{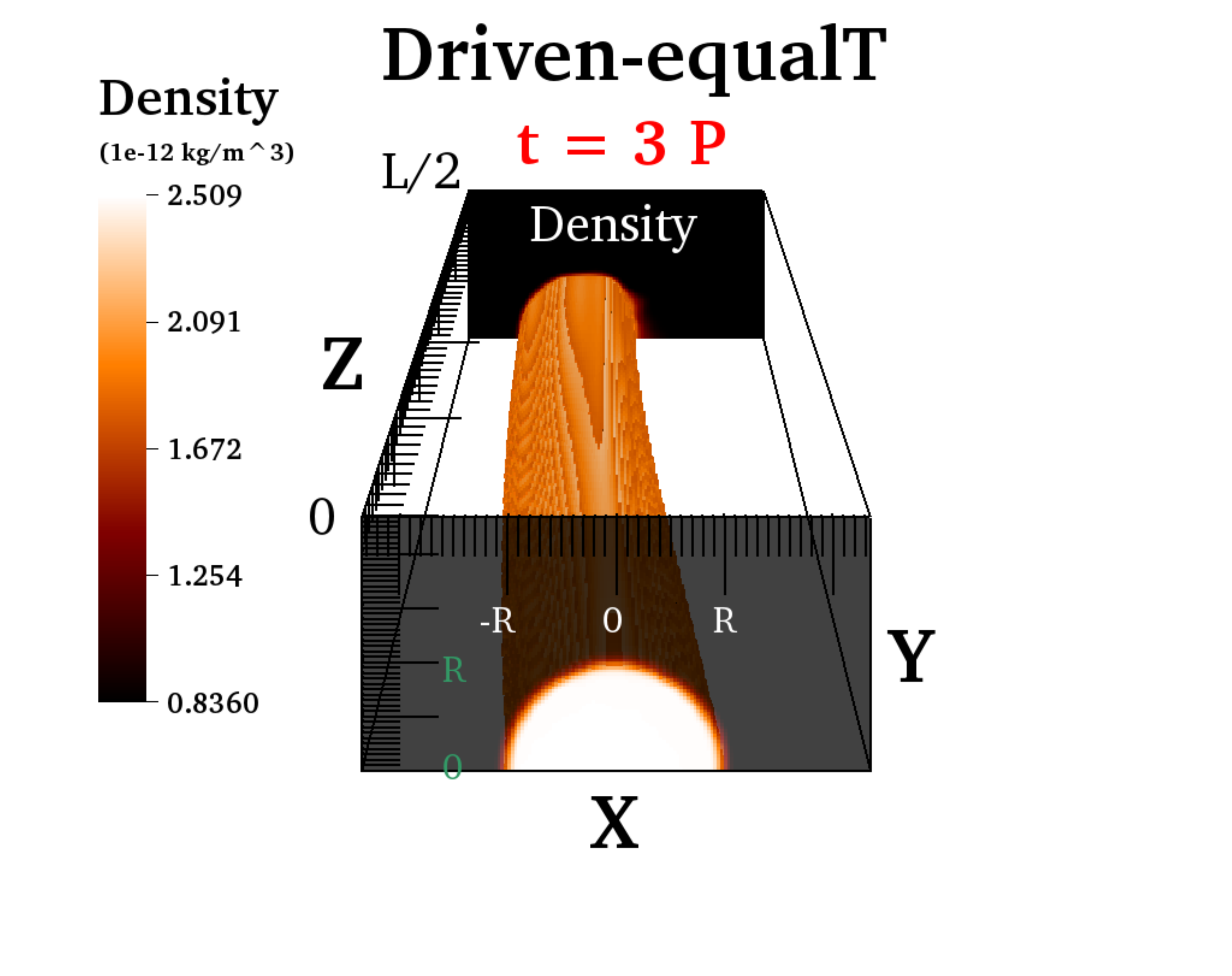}
\caption{$3$D density plot, measured in $10^{-12}$ kg/m$^3$, of our basic setup (t = 0), and of the three different models at later times. The cross sections on the $x-y$ planes at the footpoint ($z=0$) and the apex ($z=L/2=100$ Mm) are shown. The region with the highest refinement level is defined by $0 \leq z \leq 100$ Mm, $\vert x \vert \leq 2.33$ Mm and $y \leq 2.33$ Mm. Animations of these figures, showing the oscillations for the three models, are available online (Movies 1-3).}\label{fig:tubesetup}
\end{figure}

The basis of our $3D$ numerical models consists of a straight, density-enhanced magnetic flux tube, in a low-$\beta$ coronal environment (Fig. ~\ref{fig:tubesetup}). Our setup follows closely the one in \citet{antolin2014fine}, with the values of our physical parameters listed in Table ~\ref{tab:param}. The index i (e) denotes internal (external) values. 

Initially, the system is permeated by a uniform magnetic field $B_0\approx 22.8 G$ directed along the flux tube, meaning in the z direction. We take a continuous and steep radial profile for density, given by the relation:
\begin{equation}
\rho(x,y) = \rho_e + (\rho_i - \rho_e)\zeta(x,y), 
\end{equation}
\begin{equation}
\zeta(x,y) = \dfrac{1}{2}(1-\tanh((\sqrt{x^2+y^2}/R-1)\,b)),
\end{equation}
where  $\rho_e = 10^9 \mu m_p$ cm$^{-3}$ ($\mu = 0.5$ and $m_p$ is the proton mass ).
By $x$ and $y$ we denote the coordinates in the plane perpendicular to the loop axis, $z$  along its axis and $b$ sets the width of the boundary layer. In our setups, we take $b=20$, which gives us an inhomogeneous layer of width $\ell \approx 0.3 R$. We choose a density ratio of $\rho_e/\rho_i = 1/3$, which is within the range of estimated ratios, as derived from observational data in \citet{aschwanden2003VD}. This ratio is expected to lead to a fast damping rate of the kink mode through resonant absorption, and thus is suitable for rapid transfer of energy from transverse to azimuthal motions.

\begin{table}
 \caption{Values of principal physical parameters used in the simulations. The index i (e) denote internal (external) values.}\label{tab:param}
 \begin{center}
  \begin{tabular}{ l  r }
        \hline \hline
    Parameter & Value \\ 
    \hline 
    Loop length (L) & $200$ Mm  \\ 
    Loop radius (R) & $1$ Mm \\
    Loop density ($\rho_i$) & $2.509 \cdot 10^{-15}$ g/cm$^3$  \\ 
    $\rho_i$/$\rho_e$ & $3$ \\ 
    Loop temperature ($T_i$) & $9 \cdot 10^5$ K \\ 
    Magnetic field ($B_z$) & $22.8$ G \\ 
    Plasma $\beta$ & $0.018$ \\ 
        \hline 
  \end{tabular}
 \end{center}
\end{table}

The three different cases considered in the current work are:
\begin{enumerate}
\item A model of propagating waves in a loop continuously driven from the footpoint, with no temperature variation between itself and the background plasma (Driven-equalT model).
\item A model of propagating waves in a loop continuously driven from the footpoint, in hydrostatic equilibrium between itself and the background plasma, where we take a temperature ratio of $T_i /T_e =1/3$ (Driven-diffT model).
\item A model of a standing wave in a loop with an initial velocity perturbation and no temperature variation between itself and the background plasma (Stand-equalT model).
\end{enumerate}
These temperature profiles are very useful in dealing with the underlying heating mechanisms in the solar corona. By choosing a gradient of $T_i /T_e =1/3$, we are effectively modelling a coronal loop during a cooling phase, as observed for loops in thermal non-equilibrium \citep{froment2015tne, froment2017tne}, and we can directly compare to previous work dealing with the structure and observational signatures of transverse waves in coronal loops \citep{antolin2014fine,antolin2017arXiv170200775A}. Similarly, setting $T_i=T_e$ helps us identify the effects of the wave heating, no matter how subtle they might be. For the two models with no initial temperature variation ($T_i=T_e$), due to the pressure gradient between the tube and the environment, we introduced a slight decrease in the magnetic field within the tube, thus restoring total pressure equilibrium. The external Alfv\'{e}n speed for all three models is equal to $\upsilon_{Ae}=2224$ km s$^{-1}$. The internal Alfv\'{e}n speed is $\upsilon_{Ai}=1284$ km s$^{-1}$ for the model with $T_i /T_e =1/3$, and $\upsilon_{Ai}=1276$ km s$^{-1}$ for the two models with $T_i = T_e$.

\subsection{Grid}

The three dimensional ideal magnetohydrodynamic (MHD) problem is solved using the MPI-AMRVAC code \citep{keppens2012, porth2014}, where Powell’s scheme is employed to keep the solenoidal constraint on the magnetic field. We use the implemented second-order ‘onestep’ TVD method with the Roe solver and ‘Woodward’ slope limiter. Our domain dimensions, in Mm, are $(x,y,z) = ((-8,8),(0,8),(0,100))$, with four levels of refinement present, which leads to an effective resolution of $512 \times 256 \times 64$. This translates into cell dimensions of $31.25 \times 31.25 \times 1562.5$ km, thus the resolution is higher in the $x-y$ plane, to resolve the small-scale phenomena that appear around the loop edge. The loop footpoint for each model is located at $z=0$ and the apex at $z=100$ Mm. Numerical resistivity is present in our model and has a value many orders of magnitude larger than the expected one in the solar corona, which can only be reduced through the use of an ever-more refined computational grid.
Using a parameter study in all three models, we have estimated the maximum effective numerical resistivity to be of the order of $8.5 \cdot 10^{-9}$ s (in CGS). These calculations give a Lundquist number:
\begin{equation}
S = \dfrac{4\pi}{c^2} \dfrac{l \upsilon}{\eta_n} \geq 2.1 \cdot 10^4
,\end{equation}
and a corresponding resistive time scale is $\tau_{res}= 4 \pi l^2 / (c^2 \eta_n) = 1.65 \cdot 10^4$ s. Here we considered a characteristic velocity $\upsilon = 1.3 \, Mm/s \approx \upsilon_{Ai}$ and a characteristic length $l = 1$ Mm.

\begin{figure}
\resizebox{\hsize}{!}{\includegraphics[trim={1cm 0cm 1cm 1cm},clip,scale=0.27]{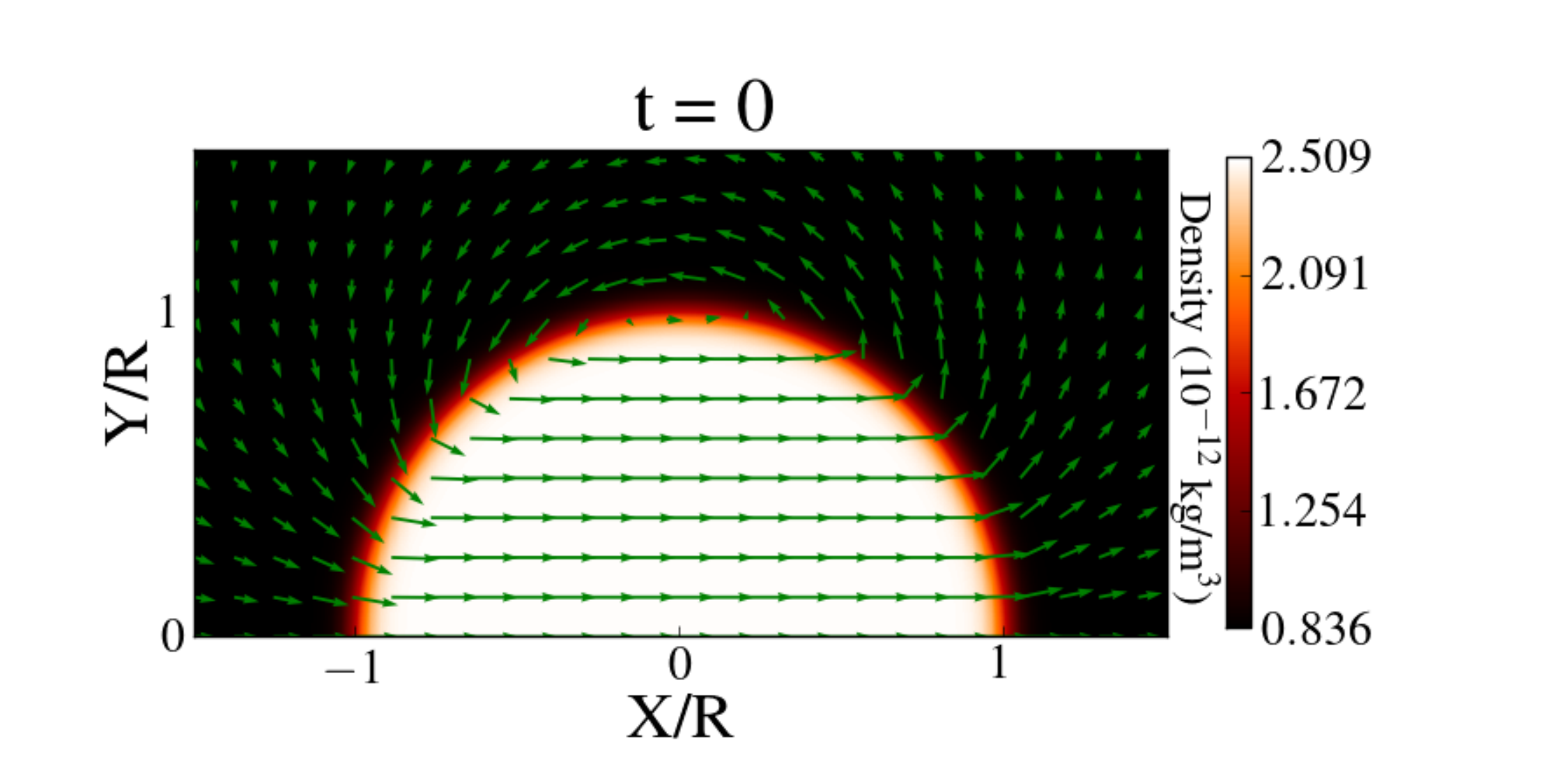}}
\resizebox{\hsize}{!}{\includegraphics[trim={1cm 0cm 1cm 1cm},clip,scale=0.27]{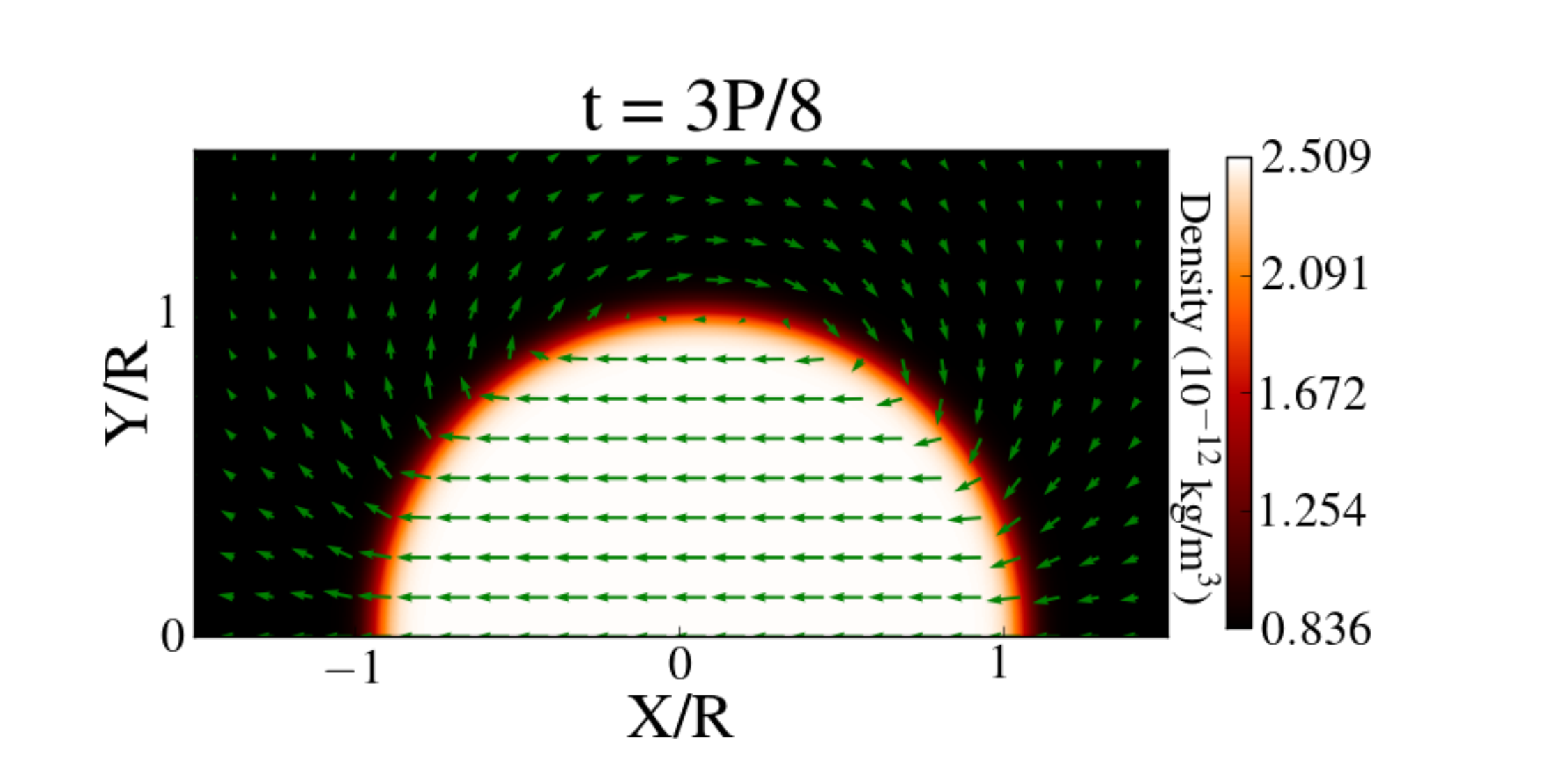}}
\caption{Contour of our tube density profile on the $x-y$ plane, at the footpoint. The vector field represents the spatial dependence of our driver for time (top) $t=0$ and (bottom) $t=3 P/8$. $P=254$ s is the driver period. The arrow lengths represent the normalized velocity, with respect to $v_0$, of our driver.}\label{fig:driver}
\end{figure}

\subsection{Driver}
Our tubes are driven from the footpoint ($z=0$ Mm), using a continuous, monoperiodic `dipole-like' driver, inspired by the one used by \citet{pascoe2010}. The period of the driver is $P\simeq 2L/c_k\simeq 254$ s for both models, coinciding with their corresponding fundamental eigenfrequency \citep{edwin1983wave}. The Stand-equalT model also has the same fundamental eigenfrequency, as the other two models. 

The driver velocity is uniform inside the loop and time varying:
\begin{equation}
\lbrace v_x,v_y \rbrace=\lbrace v(t),0 \rbrace = \lbrace v_0 \cos(\dfrac{2\pi t}{P}),0 \rbrace ,
\end{equation}
where $v_0$ is the peak velocity amplitude. Here we choose $v_0 = 2$ km/s, which is close to the observed photospheric motions. 
Outside the loop, the velocity follows the relation:
\begin{equation}
\lbrace v_x,v_y \rbrace = v(t)R^2 \lbrace \frac{x^2-y^2}{(x^2+y^2)^2},\frac{2xy}{(x^2+y^2)^2} \rbrace
.\end{equation}
To avoid any numerical instabilities due to jumps in the velocity, a transition region between the two areas exists, the shape of which matches that of the density profile. Furthermore, our driver follows the movement of the tube, ensuring that the base of the tube and only that, is always inside the central region of uniform velocity (Fig. ~\ref{fig:driver}).

For comparison, we also ran a simulation with no driver but with an initial velocity perturbation (Stand-equalT model) of the form:
\begin{equation}
\lbrace v_x,v_y,v_z \rbrace=\lbrace v(x,y,z),0,0 \rbrace = \lbrace v_0 \cos(\dfrac{\pi z}{L}) \, \zeta(x,y),0,0 \rbrace,
\end{equation}
where $v_0 = 25$ km/s. This way, the loop is subject to a perturbation mimicking a fundamental kink mode.

\subsection{Boundary conditions}
For all three models, the velocity component parallel to the $z$ axis ($v_z$) is antisymmetric at the bottom boundary in order to prevent flow of mass from the tube into, what would be, the photosphere (or `out of the loop'). The rest of the physical variables there obey a Neumann-type, zero-gradient, condition, except the $v_x$ and $v_y$ velocities for the Driven- models, which are defined by the driver. Our aim is to study the fundamental standing kink mode for an oscillating flux tube. Taking advantage of this mode inherent symmetries, as well as the symmetric nature of our driver, we simulated only one quarter of the loop (Fig. ~\ref{fig:tubesetup}). Along the axis, we went from one footpoint to the apex. We kept $v_z$, $B_x$ and $B_y$ antisymmetric, in the $x-y$ plane at the apex, while all the other quantities are symmetric. Additionally, we took into account the symmetric nature of our driver along the $y$ axis, $v_y$ and $B_y$ are antisymmetric in the $x-z$ plane, while the other quantities are symmetric. Therefore, our computational time is reduced four-fold in total for all three cases. At the three lateral boundaries, we apply outflow (Neumann-type, zero-gradient condition) conditions, which allow waves to leave the domain. To minimize their influence on the dynamics, we placed them at a safe distance from the loop ($8$ R in the $x$ and $y$ direction).

\begin{figure}
\includegraphics[trim={0.5cm 0cm 5.17cm 0cm},clip,scale=0.3]{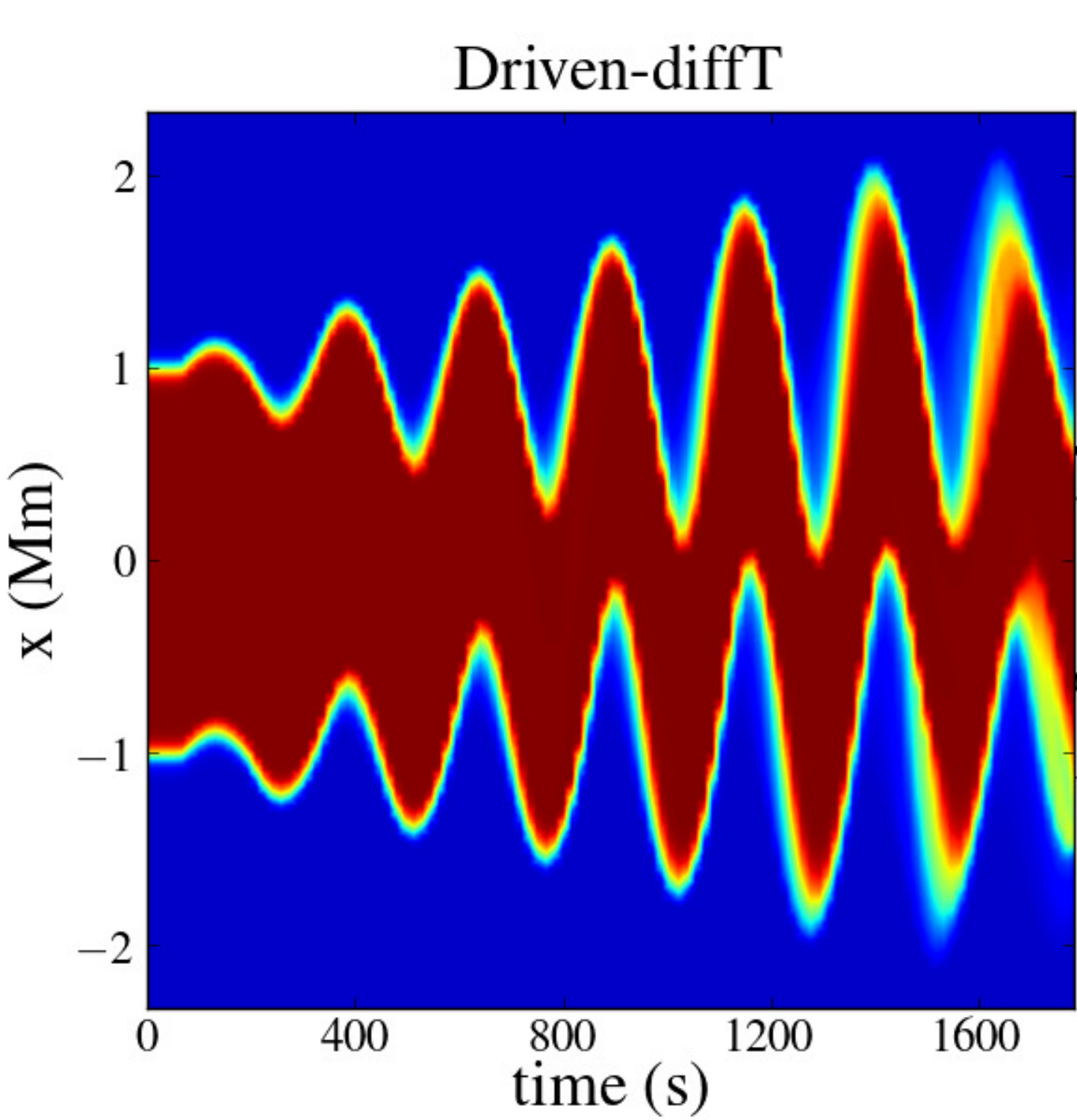}
\includegraphics[trim={2.35cm 0cm 2cm 0cm},clip,scale=0.3]{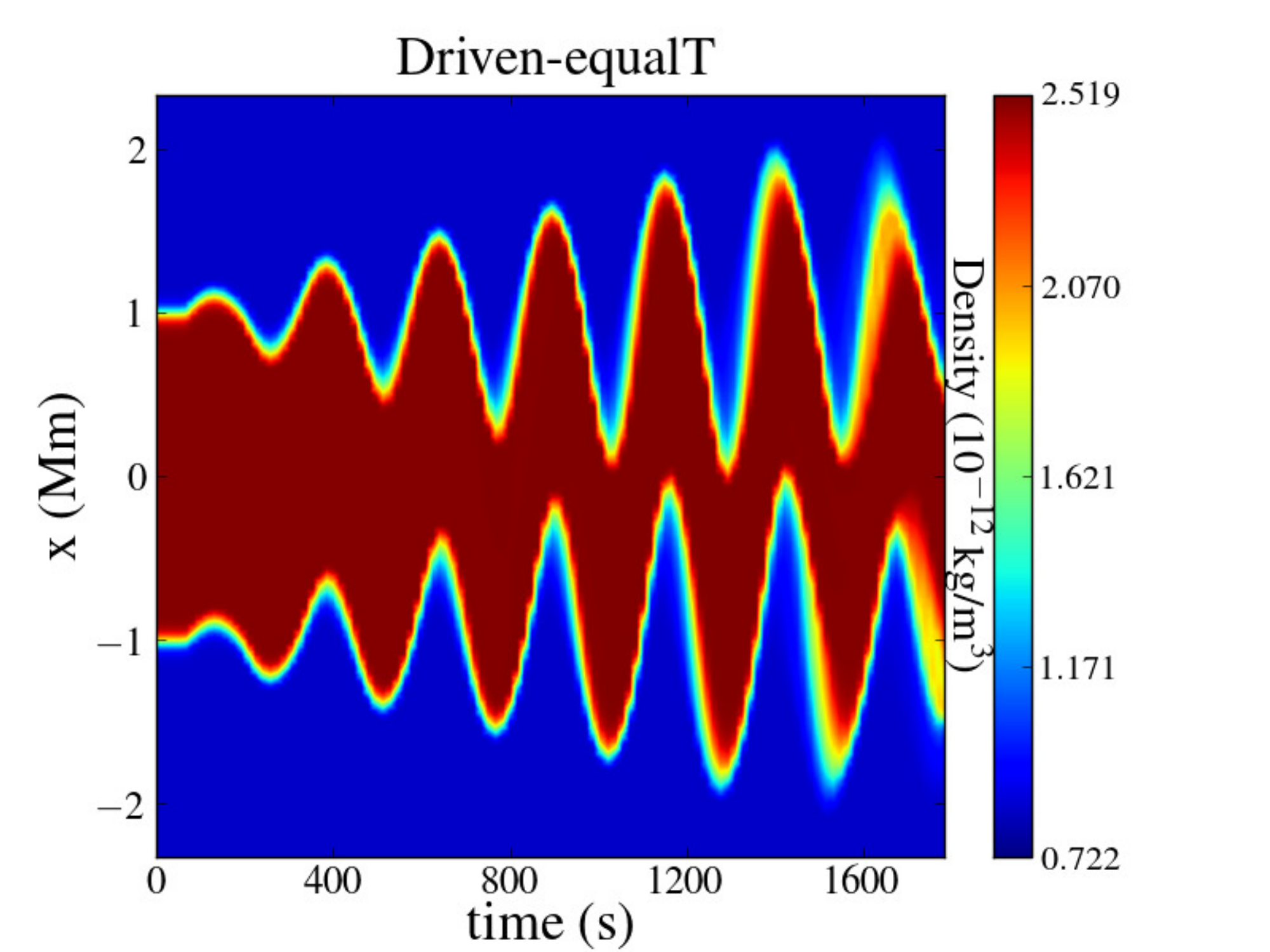}
\includegraphics[trim={0.5cm 0cm 2cm 0cm},clip,scale=0.3]{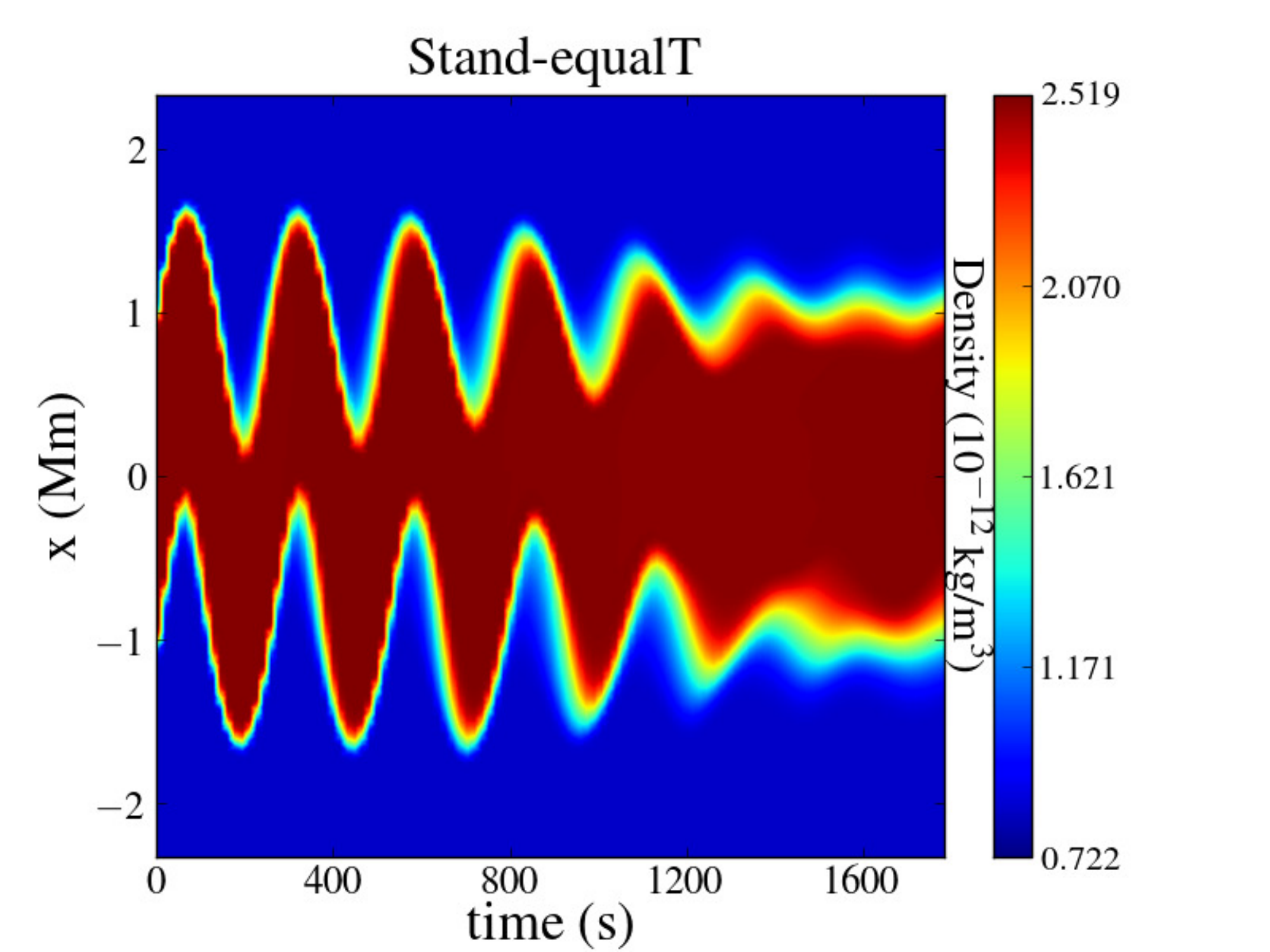}
\caption{Time-distance maps of density at the apex for: (top-left) the Driven-diffT model, (top-right) the Driven-equalT model and (bottom) the Stand-equalT model.}\label{fig:rhot}
\end{figure}

\begin{figure}
\resizebox{\hsize}{!}{\includegraphics[trim={0.1cm 0cm 1.4cm 1.3cm},clip,scale=0.23]{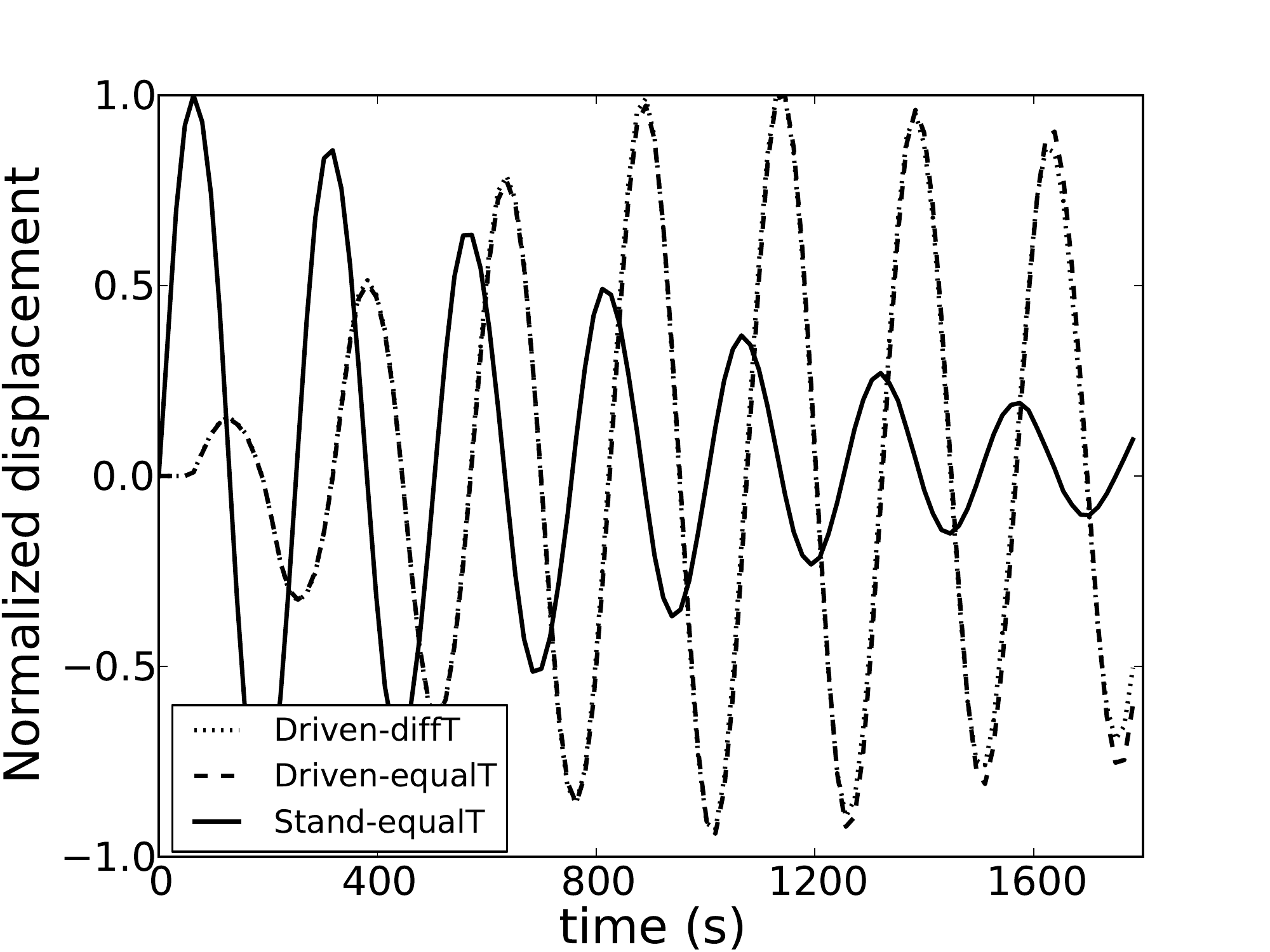}
\includegraphics[trim={0.4cm 0cm 1.4cm 1.3cm},clip,scale=0.23]{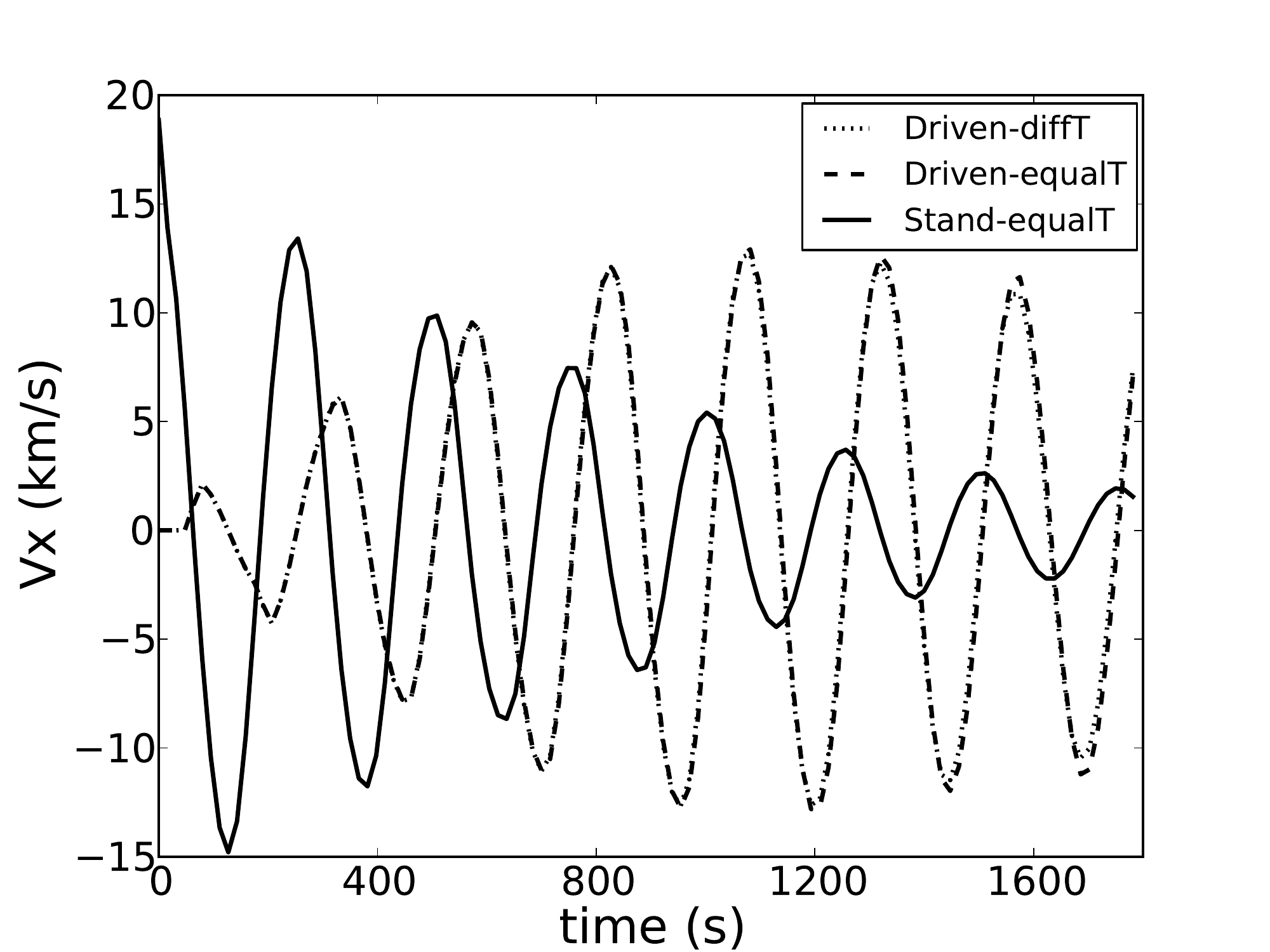}}
\caption{Left: Normalized amplitude of the loop displacement, for the three models, calculated by tracking the centre of mass at the apex ($z=100$ Mm). Right: Through centre of mass tracking at the apex, we calculate the $v_x$ velocity in km/s. The continuous line represents the tube oscillating as a standing wave with $T_e=T_i$ (Stand-equalT model), the dashed line represents the tube with the driver for $T_e=T_i$ (Driven-equalT model) and the dotted line the tube with the driver for $T_e=3\, T_i$ (Driven-diffT model).}\label{fig:cm}
\end{figure}

\section{Results and discussion}

For the rest of our analysis we focus on a sub-region of our computational domain, defined by $0 \leq z \leq 100$ Mm, $\vert x \vert \leq 2.33$ Mm and $y \leq 2.33$ Mm. This region exhibits the maximum effective resolution, and contains the loop for the whole duration of the simulation, for all three models. Inside this region, we defined the core of the loop, based on \citep{goossens2014ApJlayer}, as the part of the tube cross-section where $\rho \geq 0.976 \, \rho_i $. Furthermore, we defined the inhomogeneous layer of the tube ($0.335\, \rho_i < \rho < 0.976 \, \rho_i $), the whole tube cross-section ($ \rho > 0.335 \, \rho_i$) and, the `corona' ($\rho \leq 0.335 \, \rho_i$), all inside the same region defined previously.
 
We ran all of our simulations for a total time of seven periods ($7P \sim 1782$ s). Focusing on the driven cases, the first waves to reach the apex ($z=100$ Mm) are the azimuthal Alfv\'{e}n waves at the boundary layer of our tube, thanks to their higher propagation speed, followed by the propagating kink waves. The period of the driver is equal to the analytically predicted value for the standing fundamental kink oscillations of a uniform flux tube for our given densities \citep{edwin1983wave}. Considering the symmetry at the apex, the propagating waves from each footpoint superpose, forming a standing wave. By choosing that corresponding frequency for our driver, we forced the loop to perform an oscillation resembling the fundamental standing mode for the kink wave, with the site of the loop apex being the location of the antinode of the $x$-velocity. The animations of the tube oscillations for all three of our models are shown in the electronic version of this paper, for Fig. ~\ref{fig:tubesetup}.

In Fig.~\ref{fig:rhot}, we have the time-distance maps of the density at the apex, for our three different cases. By examining them, we can see that, for the models with the driver, the oscillation seems to saturate after reaching its peak value in amplitude, around $t \sim 1200$ s. After that point, the inner, denser part of both oscillates with a smaller amplitude, while the lower density edges maintain the maximum amplitude of the oscillation. The same drop in oscillation amplitude is present in Fig.~\ref{fig:cm}, where we plot the normalized  displacement as well as the $v_x$ velocity (given in km/s), for the centre of mass at the apex. We see that the amplitude of the oscillation, for the driven cases, reaches a maximum after $\sim 4.5$ periods (of the driver), in agreement with the time distance maps for the density. The same lower density region at the edge of the tube also develops at later times for the Stand-equalT model. The normalized display and the centre of mass velocity also reveal a damping profile, in agreement with previous works \citep{magyar2015, magyar2016damping}. 

\begin{figure*}
\centering
\includegraphics[trim={0.3cm 0.8cm 5.2cm 0cm},clip,scale=0.3]{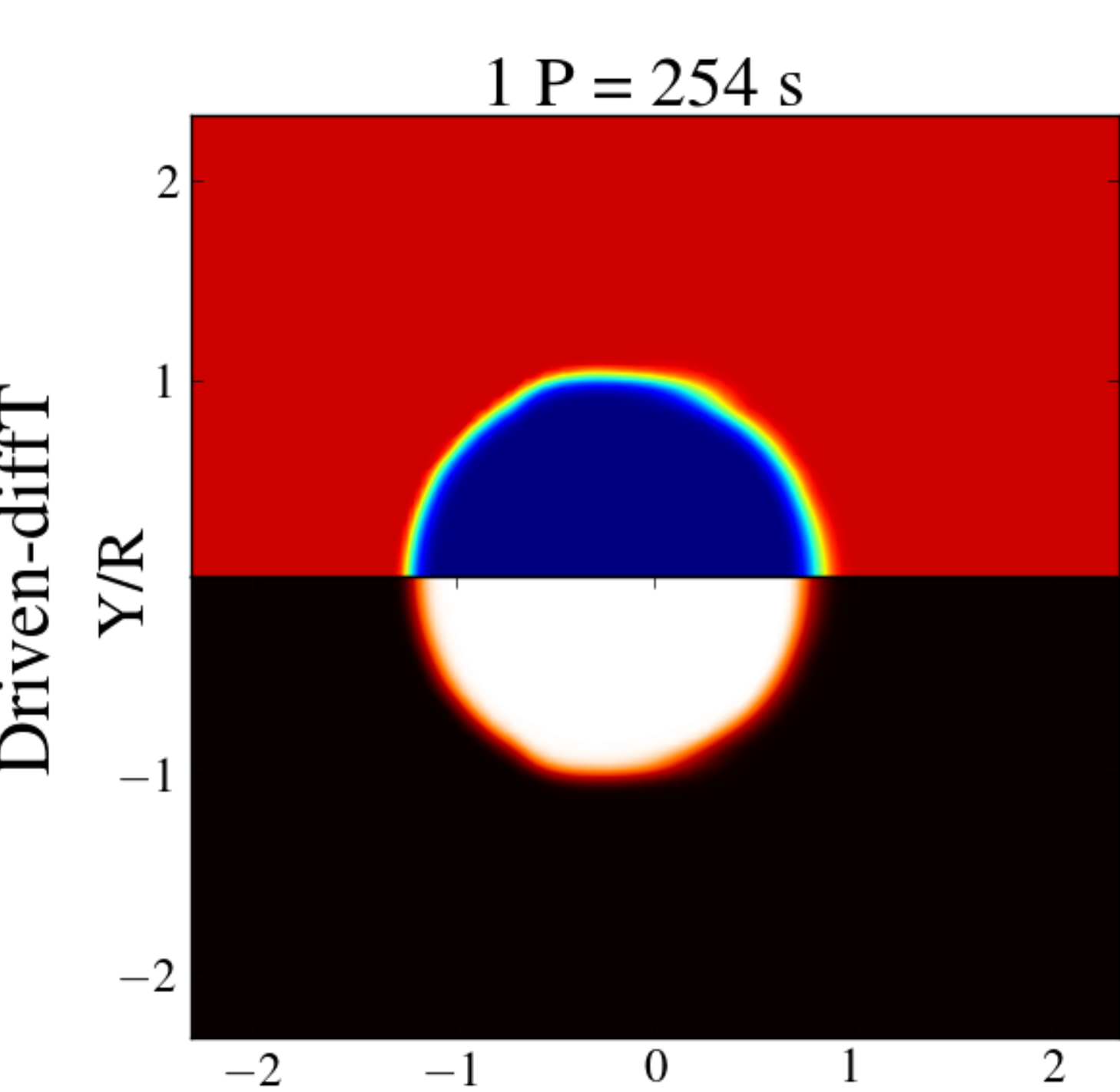}
\includegraphics[trim={2.8cm 0.8cm 5.2cm 0cm},clip,scale=0.3]{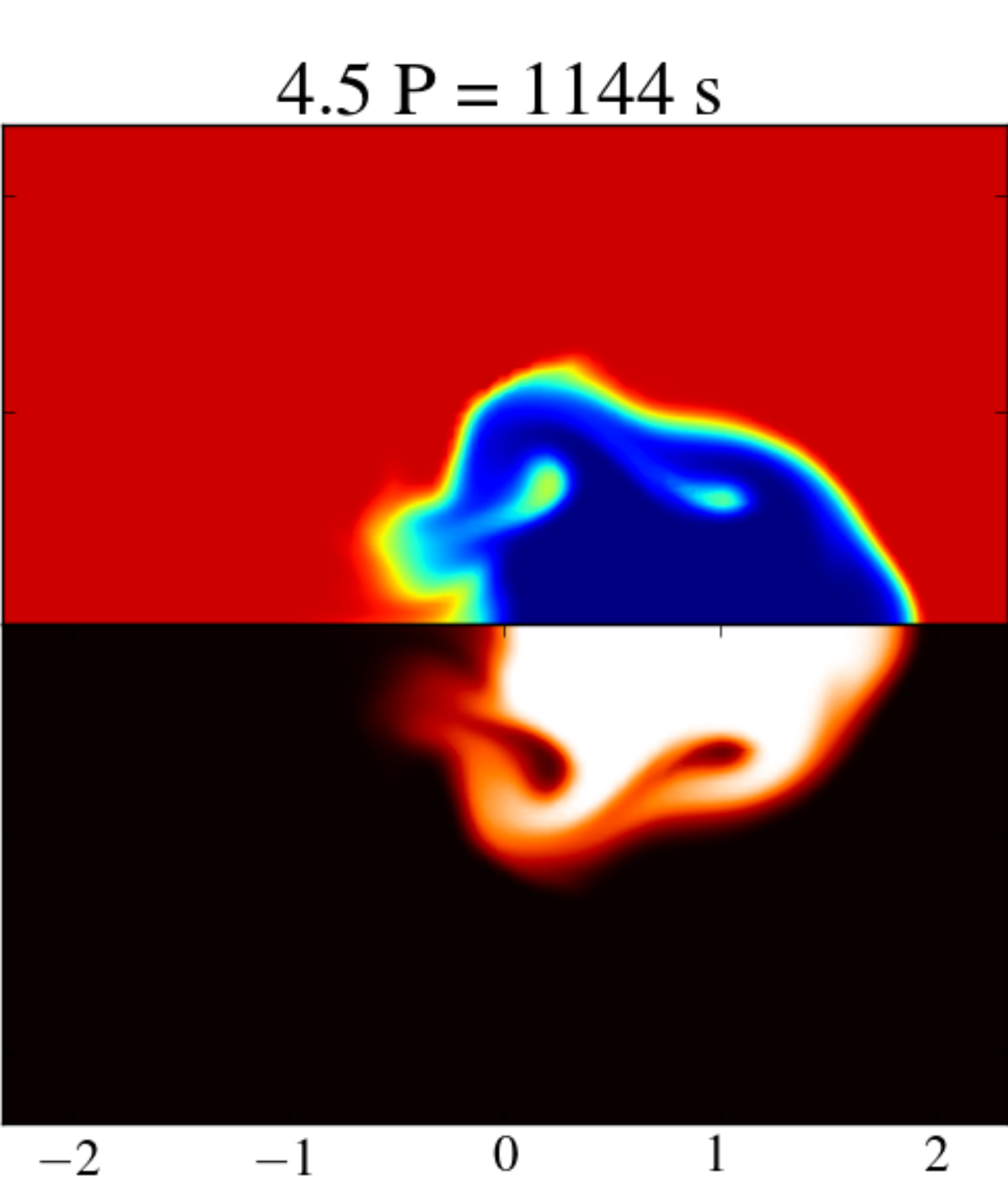}
\includegraphics[trim={2.8cm 0.8cm 2.0cm 0cm},clip,scale=0.3]{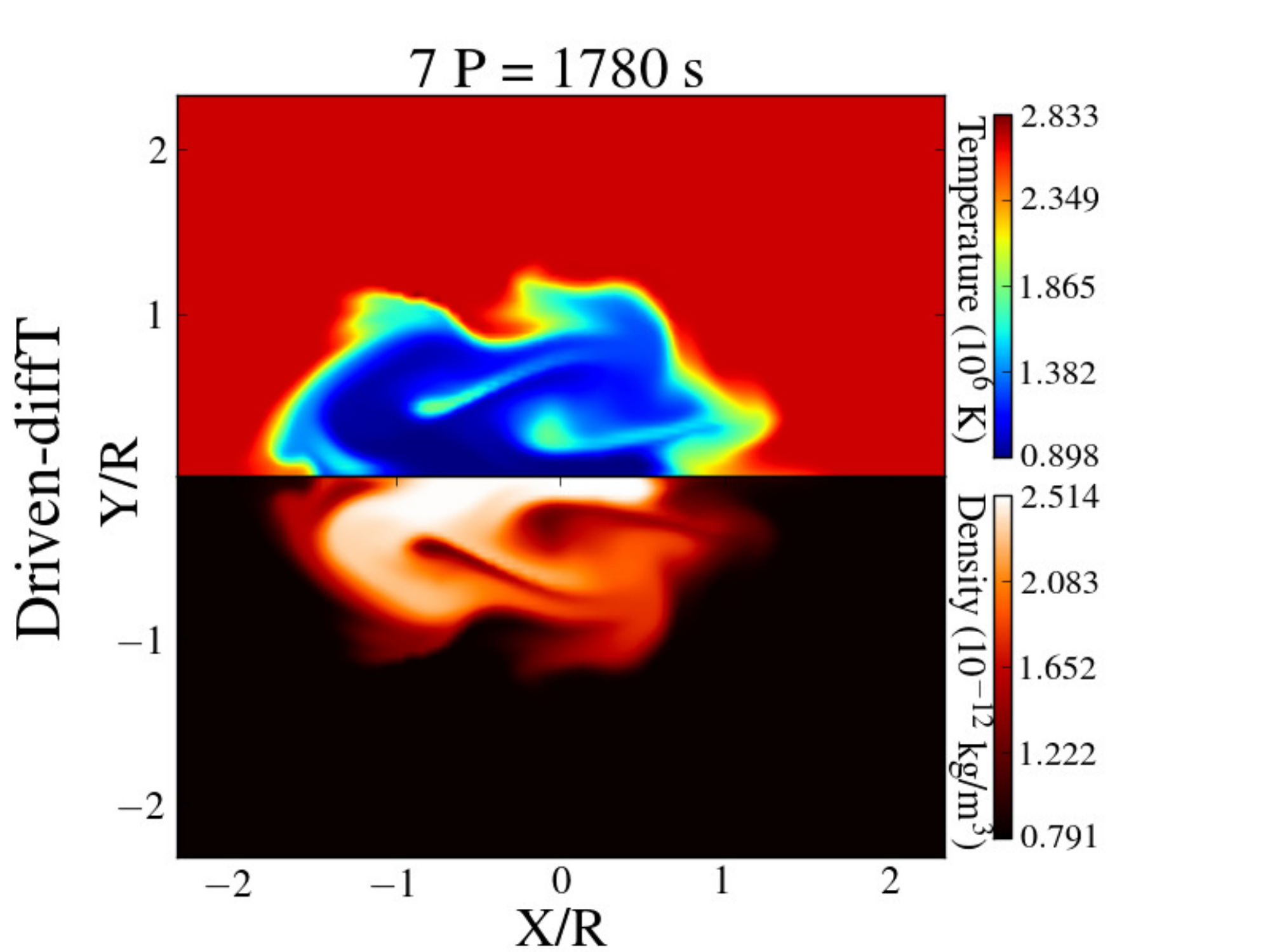}

\includegraphics[trim={0.3cm 0.8cm 5.2cm 1.5cm},clip,scale=0.3]{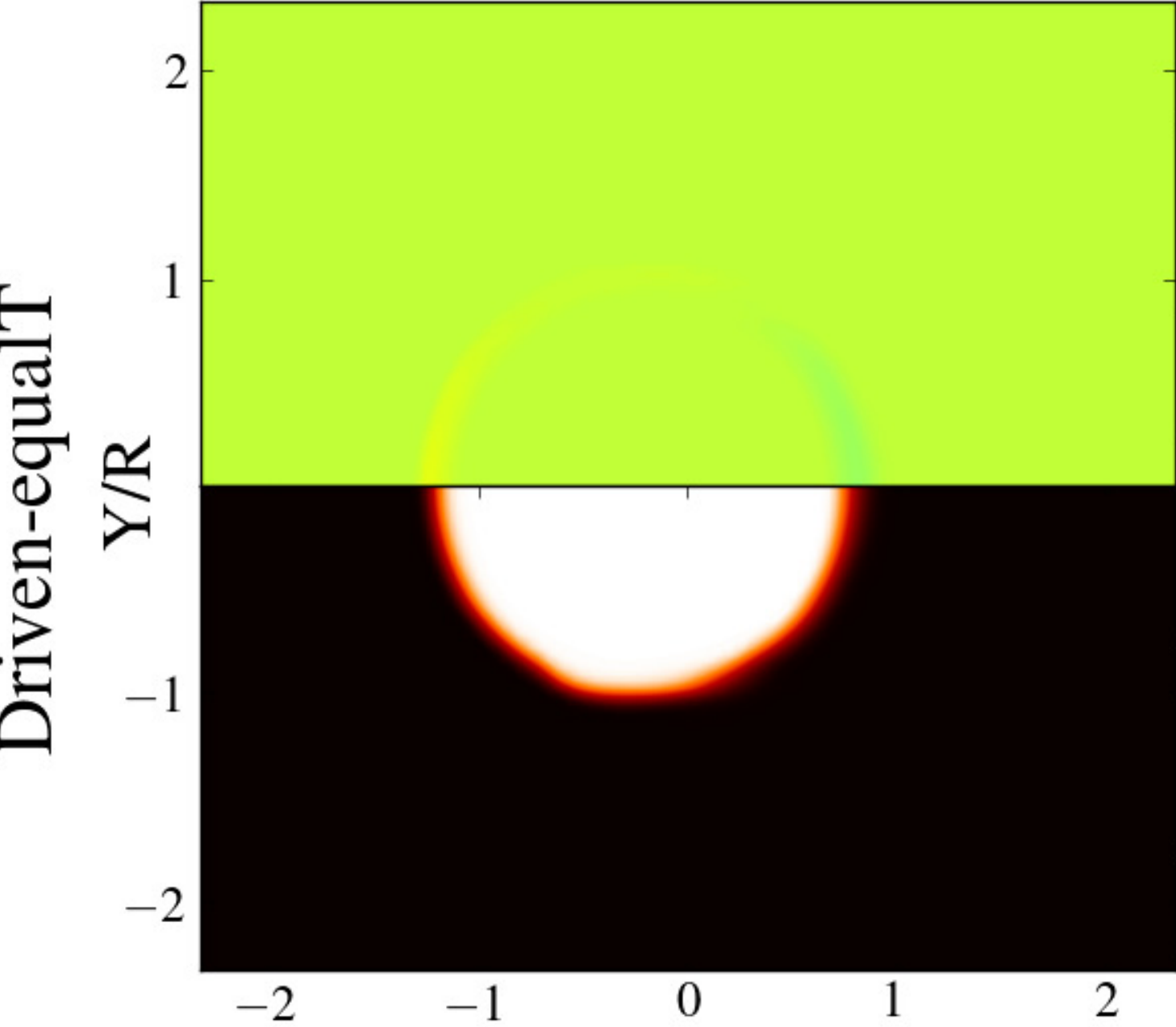}
\includegraphics[trim={2.8cm 0.8cm 5.2cm 1.5cm},clip,scale=0.3]{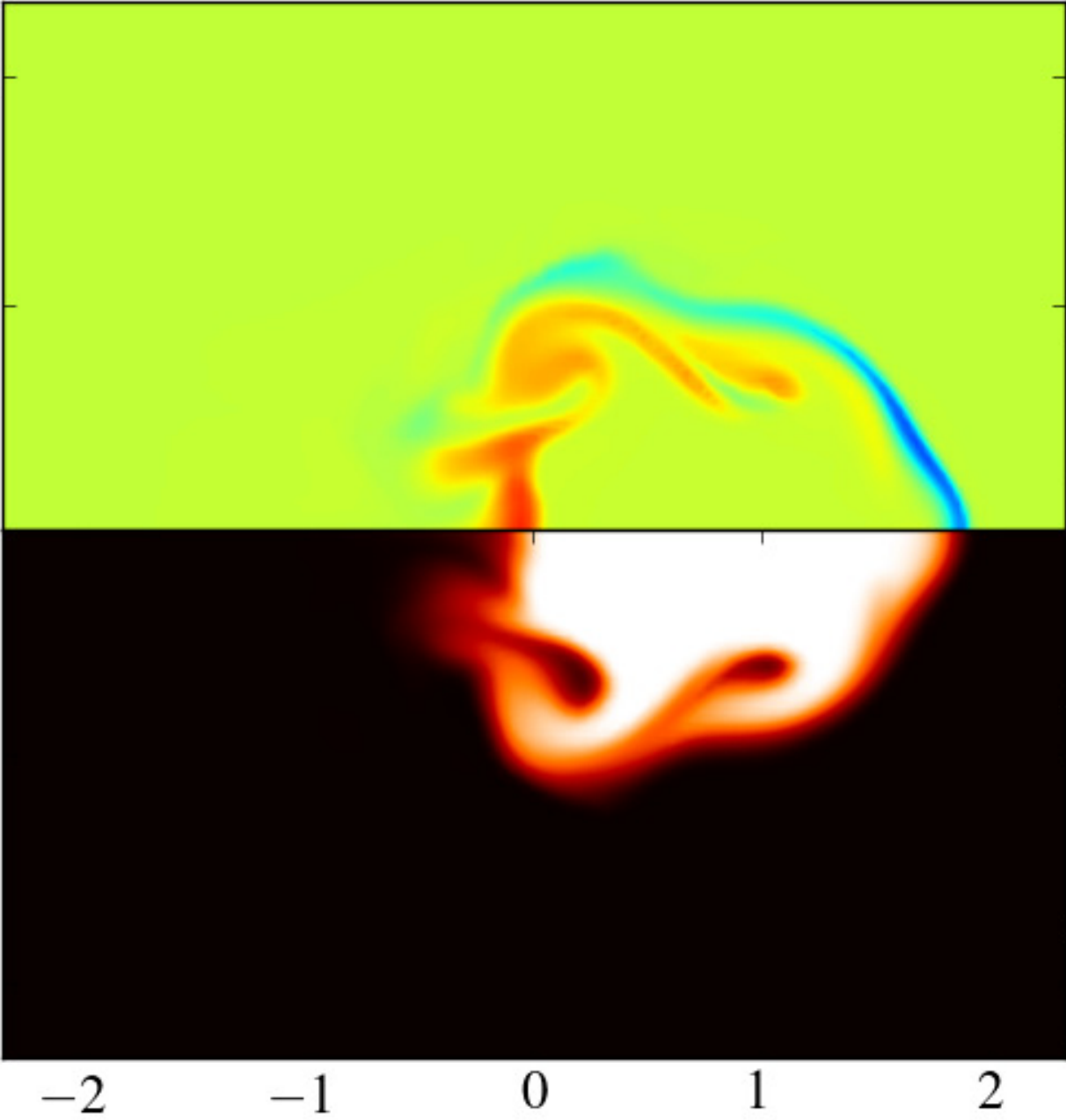}
\includegraphics[trim={2.8cm 0.8cm 2.0cm 1.5cm},clip,scale=0.3]{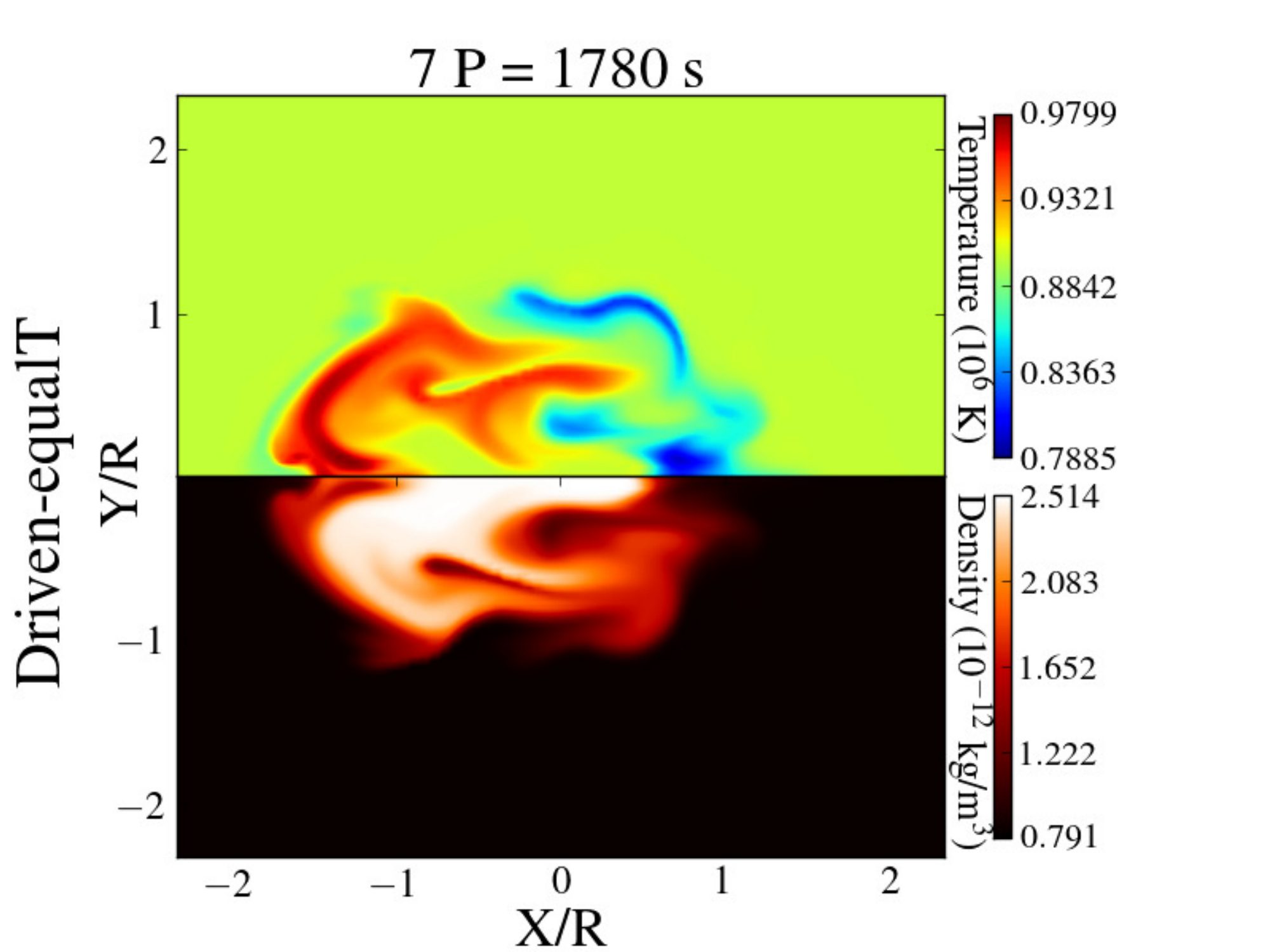}

\includegraphics[trim={0.3cm 0cm 5.2cm 1.5cm},clip,scale=0.3]{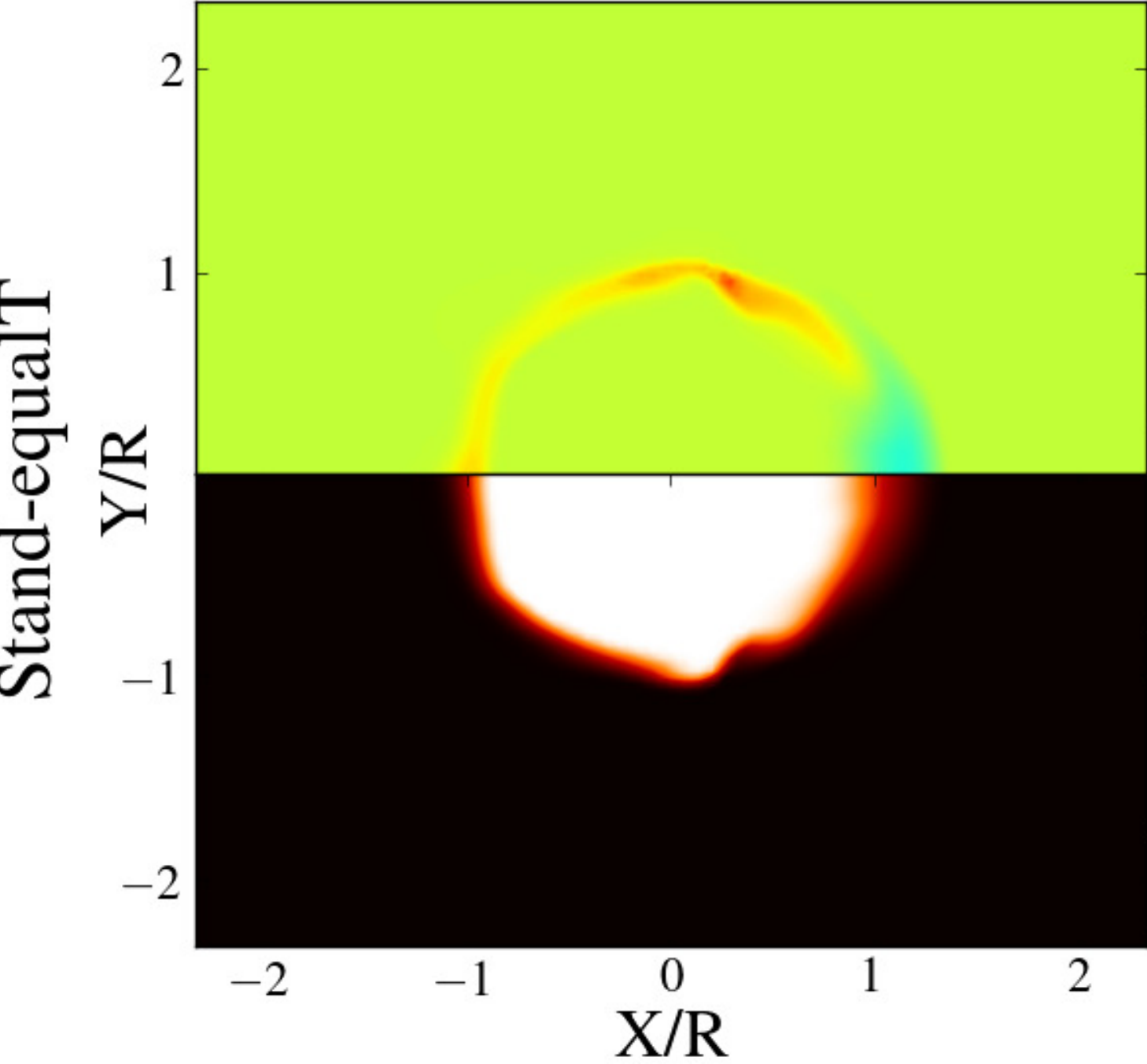}
\includegraphics[trim={2.8cm 0cm 5.2cm 1.5cm},clip,scale=0.3]{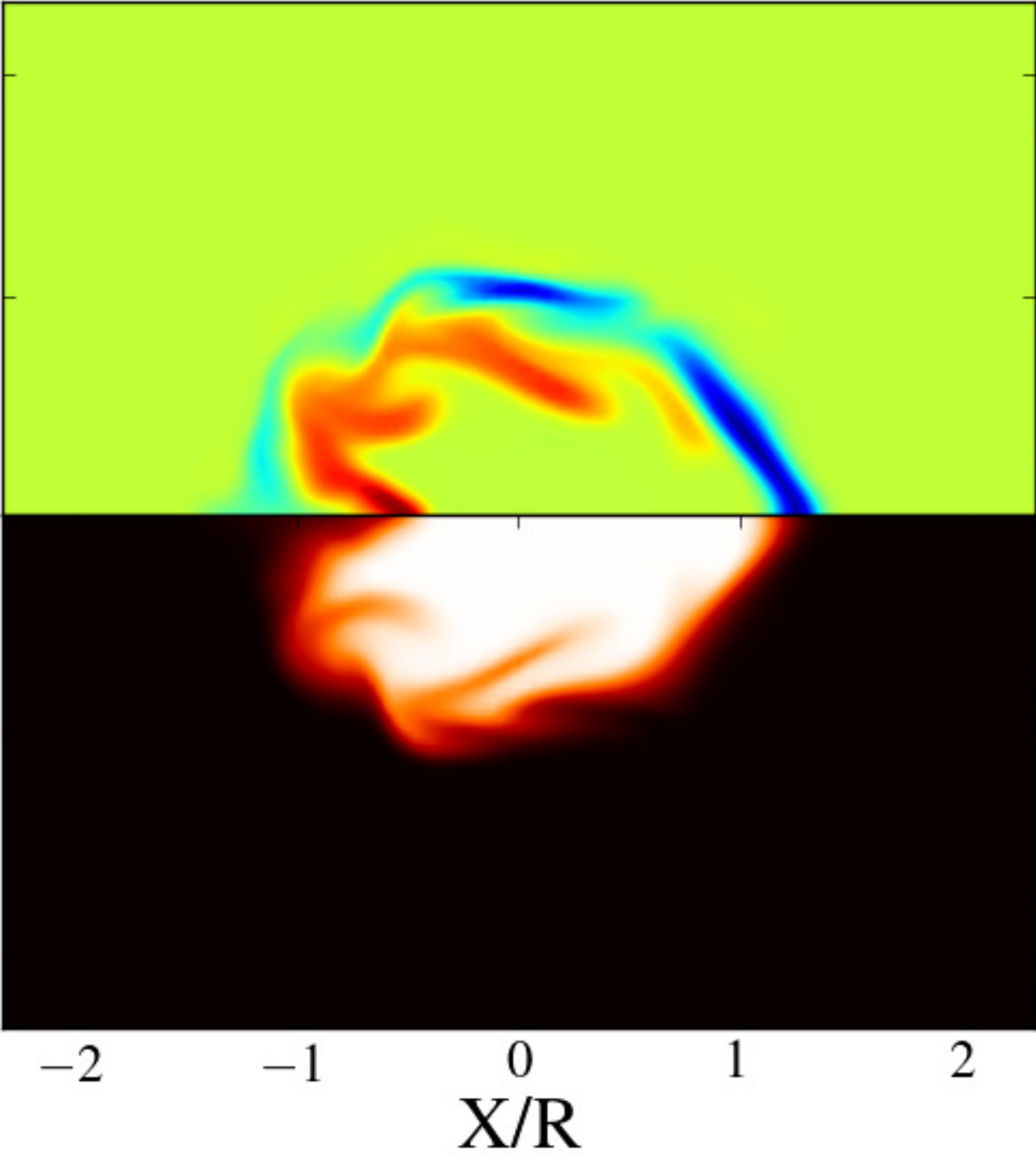}
\includegraphics[trim={2.8cm 0cm 2.0cm 1.5cm},clip,scale=0.3]{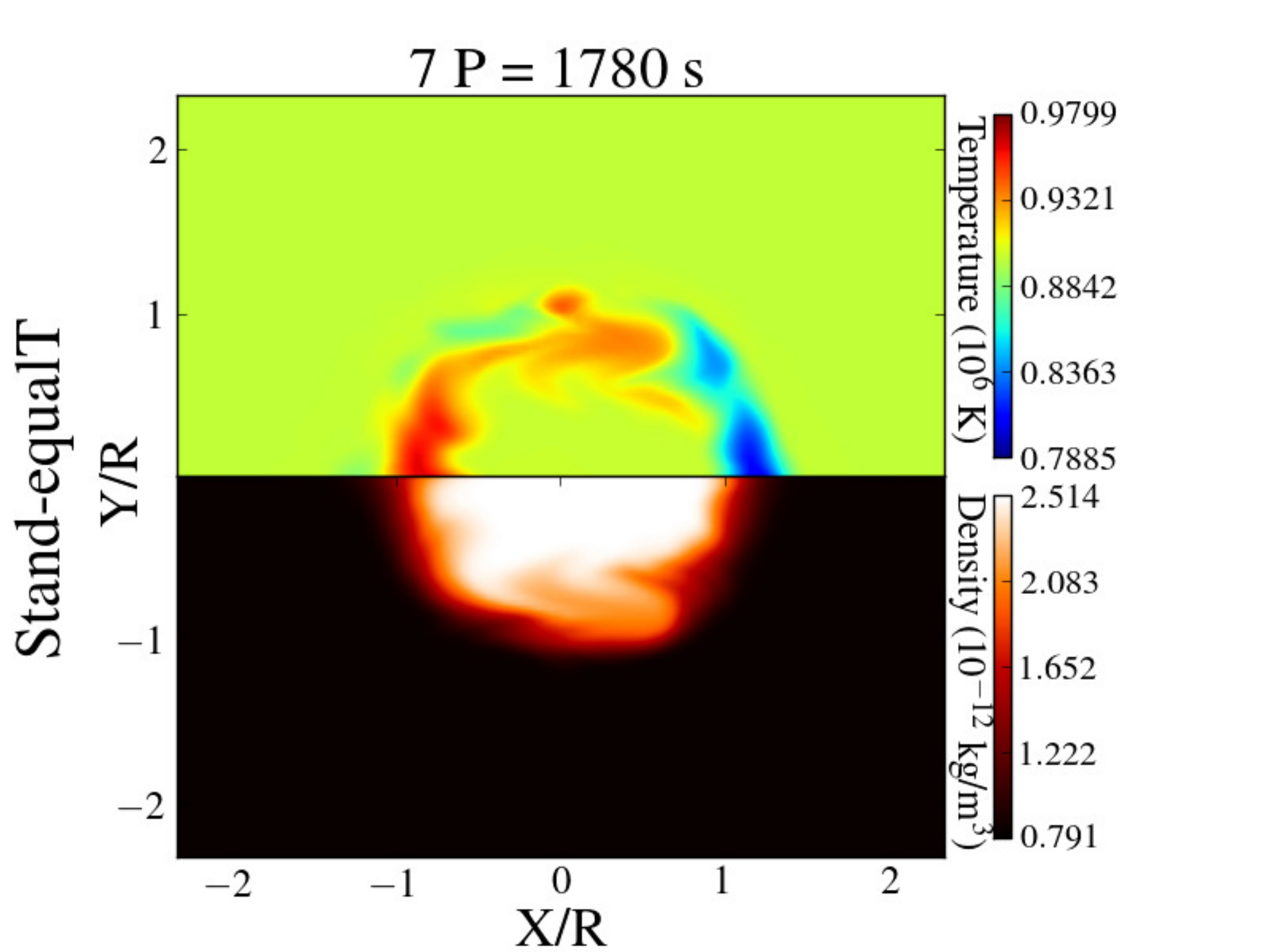}
\caption{Cross-section of the apex ($z=100$ Mm) showing the (upper half) temperature and (lower half) density profile for the three different models we studied. From top to bottom: (a) the Driven-diffT model, (b) the Driven-equalT model and (c) the Stand-equalT model.}\label{fig:TRHO}
\end{figure*}

As mentioned before, the superposing propagating waves quickly form a footpoint driven standing wave with a velocity antinode at the apex, which can be Kelvin-Helmholtz unstable \citep{heyvaerts1983,zaqarashvili2015ApJ}. In fact, approximately two periods time after the apex started to oscillate, the Kelvin-Helmholtz instability (KHI) manifests there because of the high shear velocities \citep{terradas2008,antolin2014fine, antolin2015resonant, antolin2016,magyar2015, magyar2016damping}. In Fig.~\ref{fig:TRHO}, we plot the spatial profiles of the temperature and density for our three models, at the apex. The KHI develops, creating strong shear flows and smaller scales, in addition to spatially extended eddies, the Transverse Waves Induced Kelvin-Helmholtz (TWIKH) rolls. These TWIKH rolls result in extensive mixing of plasma from the loop, with the surrounding corona, as indicated by the profiles of temperature ($T$) and density ($\rho$) at the apex. This extended turbulent layer is the low density tube edge, which we saw developing in Fig. \ref{fig:rhot}.

Focusing on the Driven-equalT and Stand-equalT models, we observe the manifestation of temperature fluctuations at the tube layer. These perturbations appear immediately after the tubes are set in motion, and are getting stronger once the Kelvin-Helmholtz instability sets in, as seen in Fig. \ref{fig:TRHO}. They are connected to the density and pressure fluctuations and are not the result of energy transfer between those different regions. \citet{antolin2017arXiv170200775A} also observed these temperature perturbations for a flux tube oscillating as a standing wave, and characterized them adiabatic in nature.  Thus, we refer to this mechanism as adiabatic heating (and cooling). Here, we prove that similar patterns for adiabatic cooling and heating appear, both for a tube with a standing oscillation and for a footpoint driven one. Adiabatic in nature temperature fluctuations can also be observed at the footpoints of our models as well, but they are more uniform and extended, probably originating from the large scale loop dynamics caused by the standing (or standing-like) oscillations. These adiabatic processes are existing alongside other heating mechanisms, which we are studying in the current work. 

\begin{figure*}
\centering
\resizebox{\hsize}{!}{\includegraphics[trim={0cm 0cm 1cm 0cm},clip,scale=0.3]{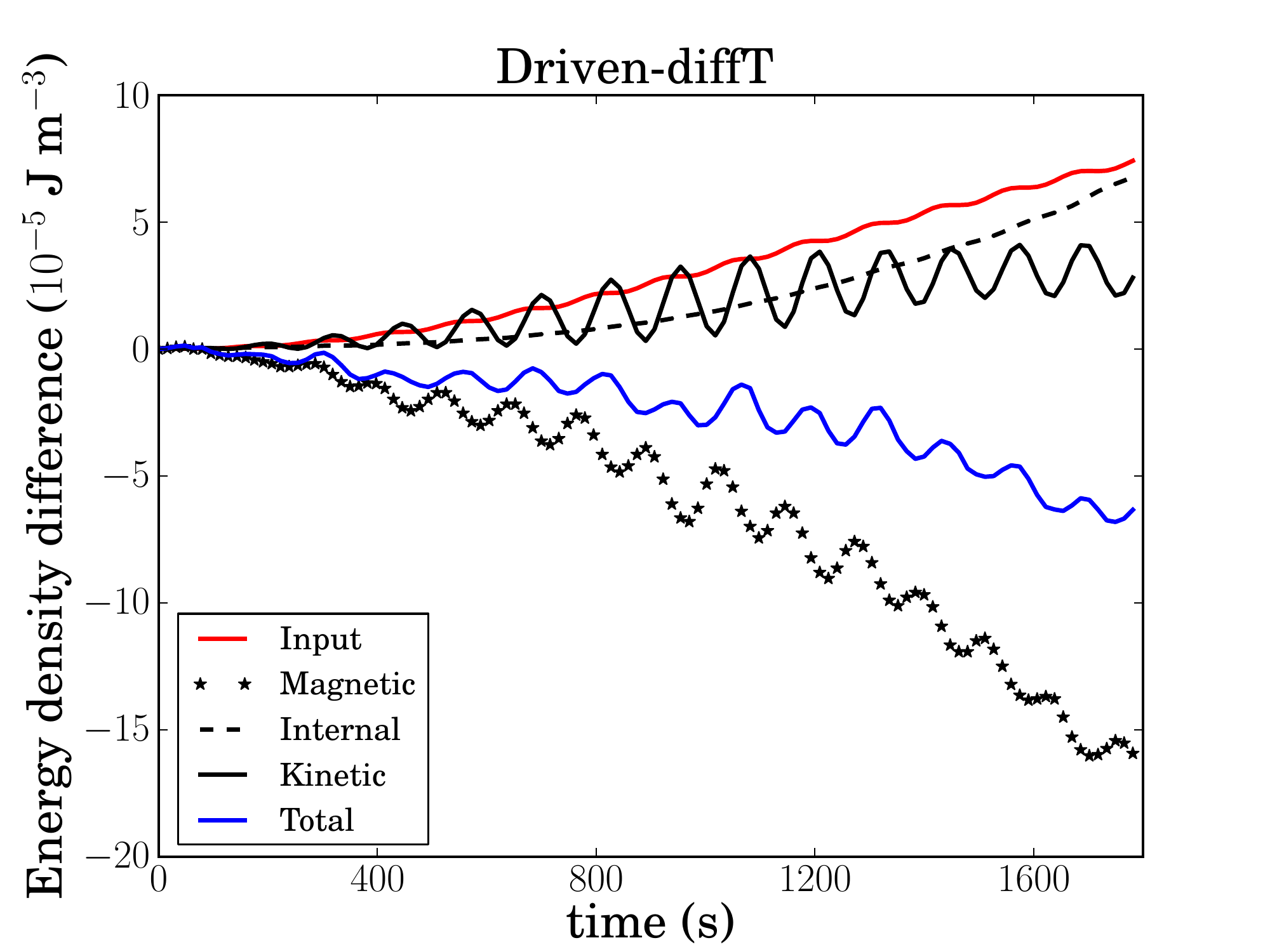}
\includegraphics[trim={0cm 0cm 1cm 0cm},clip,scale=0.3]{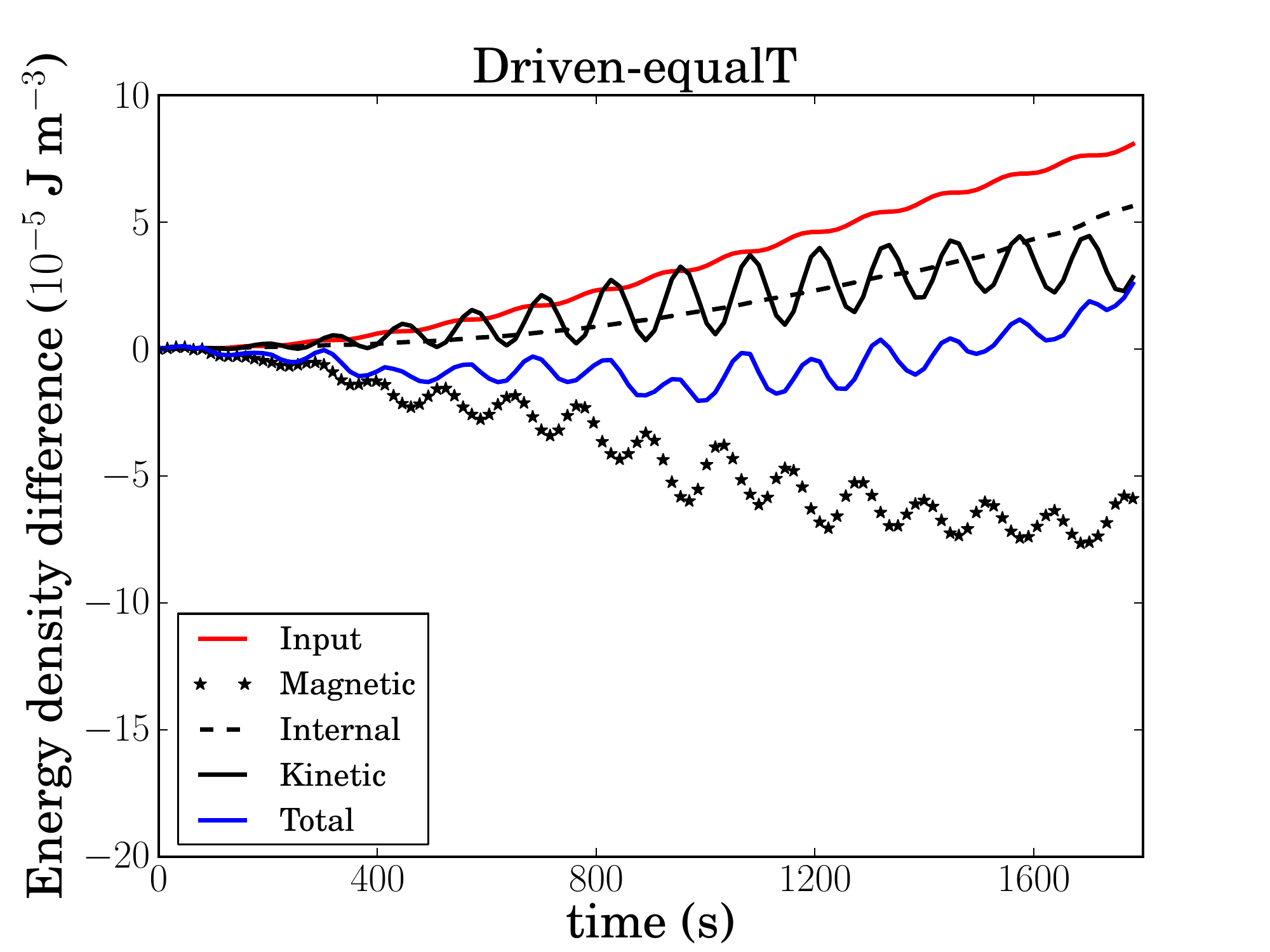}
\includegraphics[trim={0cm 0cm 1cm 0cm},clip,scale=0.3]{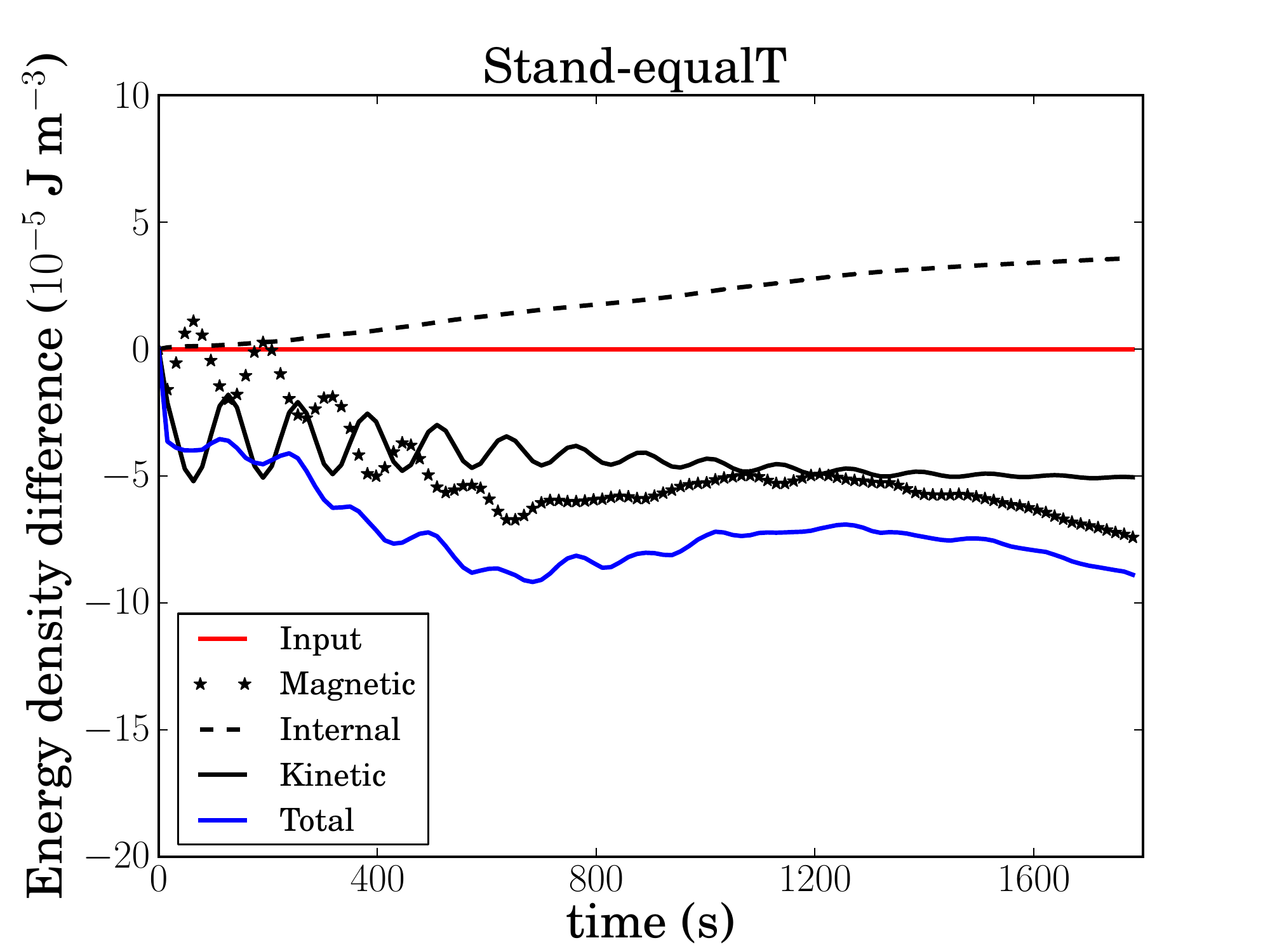}}
\caption{Time profiles for the internal, magnetic and kinetic energy density variations relative to the initial state, the total (internal$+$magnetic$+$kinetic) energy density difference and the energy density provided by the driver. All the quantities are volume averaged for the region with $0 \leq z \leq 100 $ Mm, $\vert x \vert \leq 2.33$ Mm and $y \leq 2.33$ Mm. From left to right: (a) the Driven-diffT model, (b) the Driven-equalT model and (c) the Stand-equalT model.}\label{fig:energies}
\end{figure*}

In order to investigate the drop of the normalized driven oscillation amplitudes, in Fig. \ref{fig:energies}
we plot the perturbations in internal and magnetic energy densities, the kinetic energy density and the total input energy densities for our models. Focusing on a region of constant volume, the changes in the energy densities are directly translated into changes in the energies. The total energy density provided by the driver, following \citet{belien1999ApJ}, is calculated from the formula:
\begin{equation}
S(t) = -\dfrac{1}{V}\int_0^t \int_A \textbf{S} \cdot d\textbf{A}dt',
\end{equation}
where $\textbf{S}$ is the Poynting flux (in J m$^{-2} s^{-1}$) from the lower boundary ($x-y$ plane), where the driver is located. $\textbf{A}$ denotes the surface element (of the lower $x-y$ boundary plane) and $V$ is the total volume of the studied region. As we see from the diagrams of energy, the input in the Driven-diffT and Driven-equalT models are very similar, as it was expected. The kinetic energy density shows a saturation in both cases after a time around $1200-1300$ s, that is in agreement with what we saw in the time density maps for the density. The fact that the kinetic energy saturates, indicates that the previously mentioned drop in the normalized amplitude (see Fig. \ref{fig:cm}) is not caused by an actual drop of the oscillation amplitude, but rather by the development of smaller scale motions in the loop cross-section, which affect the position of the centre of mass.   

It is also interesting to note here that the input energy, while sufficient to explain the rise of the kinetic energy, is less than the sum of internal and kinetic energy. The extra energy seems to be provided by the drop of the magnetic energy, due to the existence of the effective numerical resistivity. The same drop is higher in the case of the Driven-diffT model, but it does not drastically increase the internal energy for that model. As result, both models with the footpoint driver are expected to have very similar dynamical evolution over time. For the Stand-equalT model, the input is practically zero, as expected, and both the kinetic and magnetic energies decrease in time. 

We note here the different behaviour of the magnetic energy density difference between the three models. In all three models, not all of the available magnetic energy density turns into internal energy. From the equation for the energy density evolution in resistive MHD,  we see the existence of a resistive source term, and of fluxes. The resistive term is the one responsible for transforming the magnetic energy into internal energy, while the fluxes transfer energy into (`Input' in Fig. \ref{fig:energies}) and out of our domain. The Poynting flux through the side boundaries, which are simulated as open, is responsible for the extra drop of the magnetic energy.

For the cases of uniform temperature, only a small part of the magnetic energy density is dissipated this way, as we can see by comparing the its drop to the rise of the internal energy density. For these two cases, the magnetic energy density drops at almost the same levels, due to the identical initial conditions for the magnetic fields and the plasma pressure in these two models. The oscillatory behaviour of the Stand-equalT model magnetic energy density is caused by the initiation of the slow wave, which was mentioned before. The small rise observed near the end of the simulation for the magnetic energy of the Driven-equalT model is caused by the continuous energy input from the driver. However, the Driven-diffT model exhibits a greater drop, when compared to the other two cases. This drop cannot be adequately explained by the bigger rise of internal energy for that model (Fig. \ref{fig:energies}), and is caused by the stronger Poynting flux through the side boundaries. Finally, we need to stress that the effects of numerical resistivity, as well as the inevitable development of non-zero $\nabla \cdot B$ were taken into account during our analysis.

\begin{figure}
\resizebox{\hsize}{!}{\includegraphics[trim={0cm 0cm 1.5cm 0cm},clip,scale=0.22]{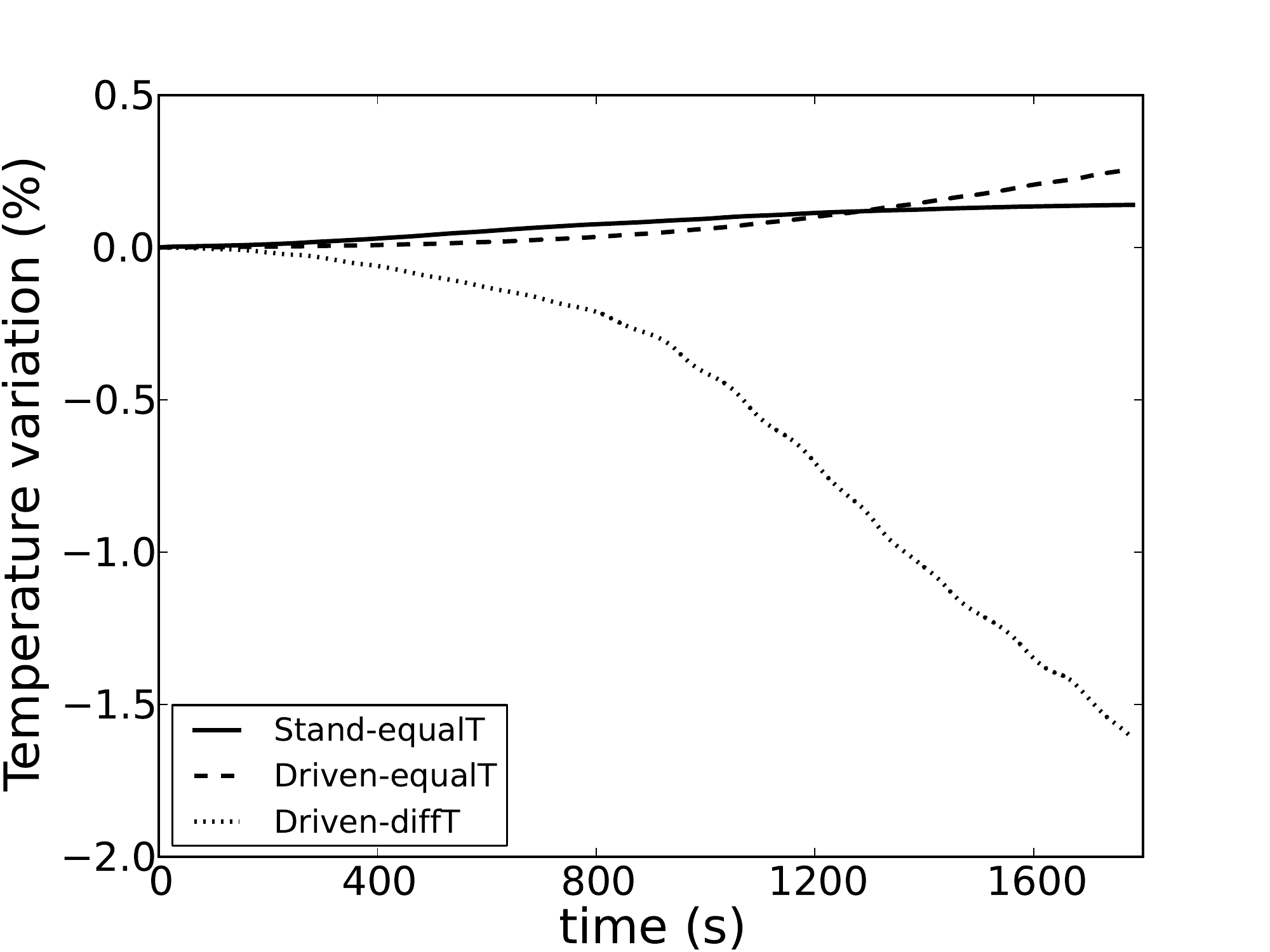}
\includegraphics[trim={0cm 0cm 1.5cm 0cm},clip,scale=0.22]{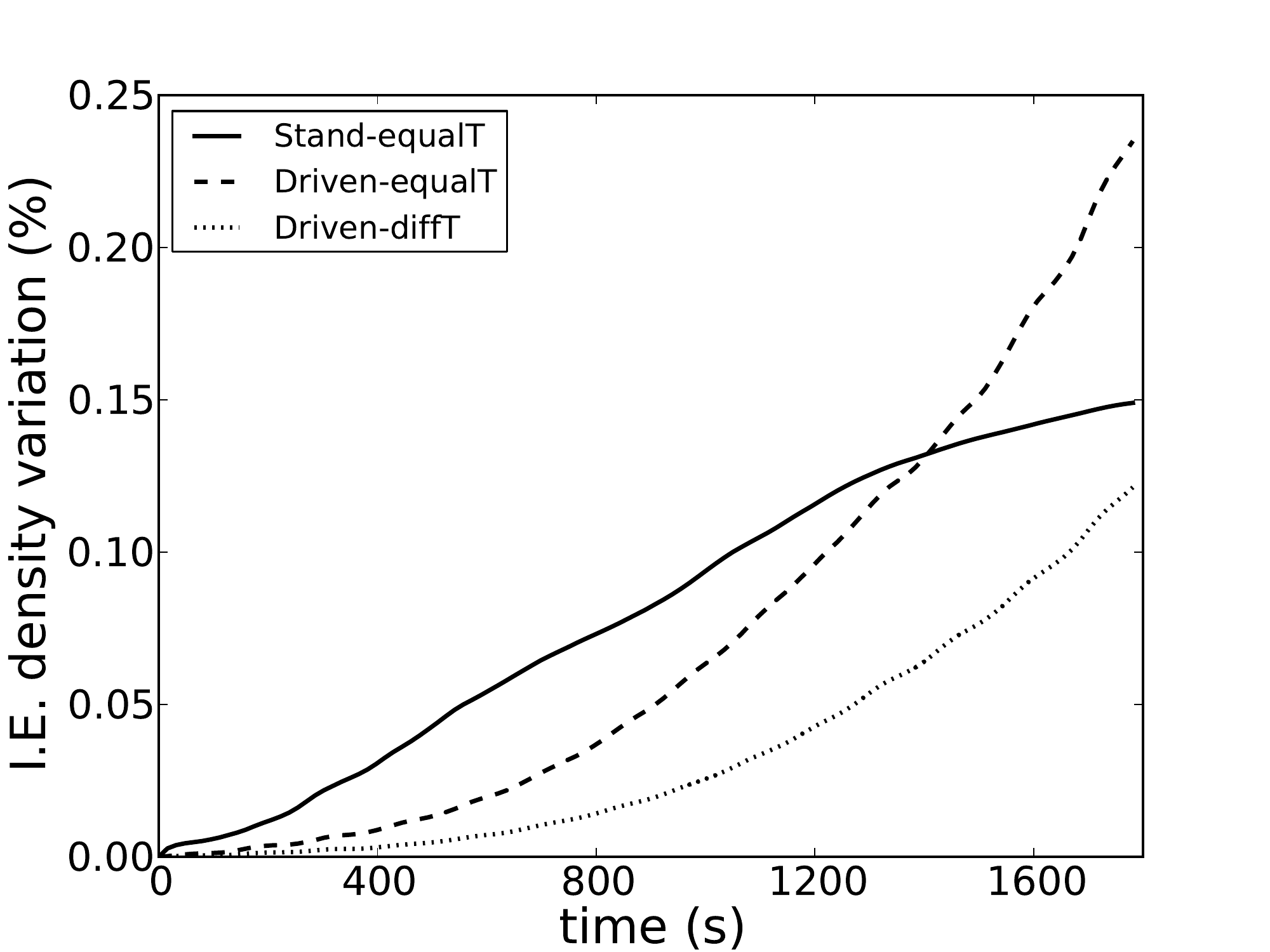}}
\caption{Percentage of the: (left) volume averaged temperature variation and (right) volume averaged internal energy variation, over a greater region (tube+corona) including the loop ($0 \leq z \leq 100 $ Mm, $\vert x \vert \leq 2.33$ Mm and $y \leq 2.33$ Mm). The continuous line represents the Stand-equalT model, the dashed line represents the Driven-equalT model and the dotted line the Driven-diffT model.}\label{fig:ietemps}
\end{figure}

In Fig. ~\ref{fig:ietemps} we plot the difference of the volume averaged temperature and of the volume averaged internal energy relative to the initial state, over the area defined by $0 \leq z \leq 100$ Mm, $\vert x \vert \leq 2.33$ Mm and $y \leq 2.33$ Mm, for all three models. We see that for the Driven-equalT and Stand-equalT models, both the temperature and internal energy show the same relative increase. This is consequence of the heating mechanisms present in our simulations, which we are going to further study in the next figures. In the case for the Driven-diffT model, the temperature exhibits a drop larger than $1.5 \%$ while the internal energy density shows a rise of about $0.12 \%$. As we explain later, this is due to the mixing between the cold tube and the hotter environment, that we considered in that particular model. 

\begin{figure*}
\centering
\includegraphics[trim={1.0cm 0.9cm 0cm 0cm},clip,scale=0.3]{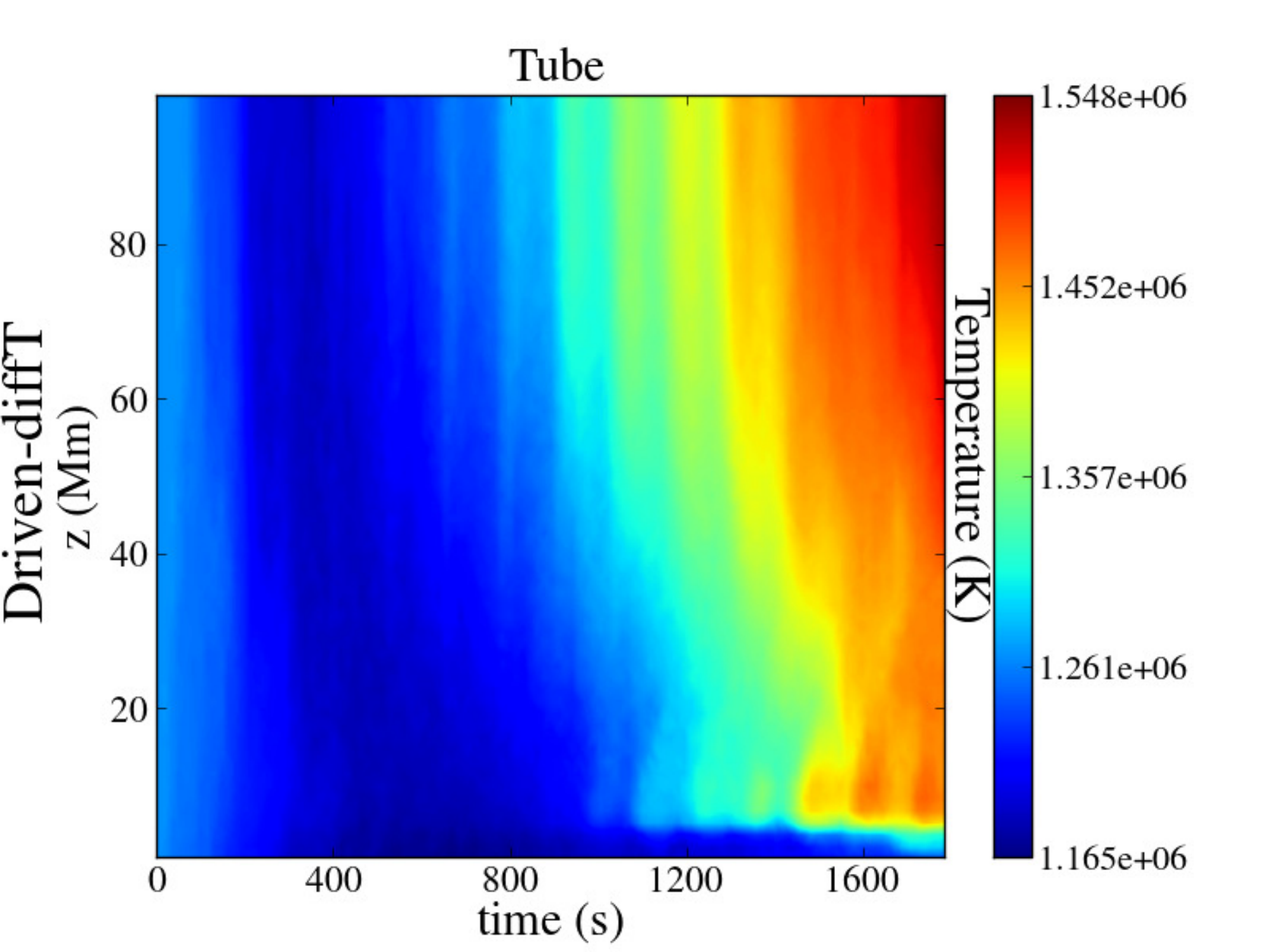}
\includegraphics[trim={1.0cm 0.9cm 0cm 0cm},clip,scale=0.3]{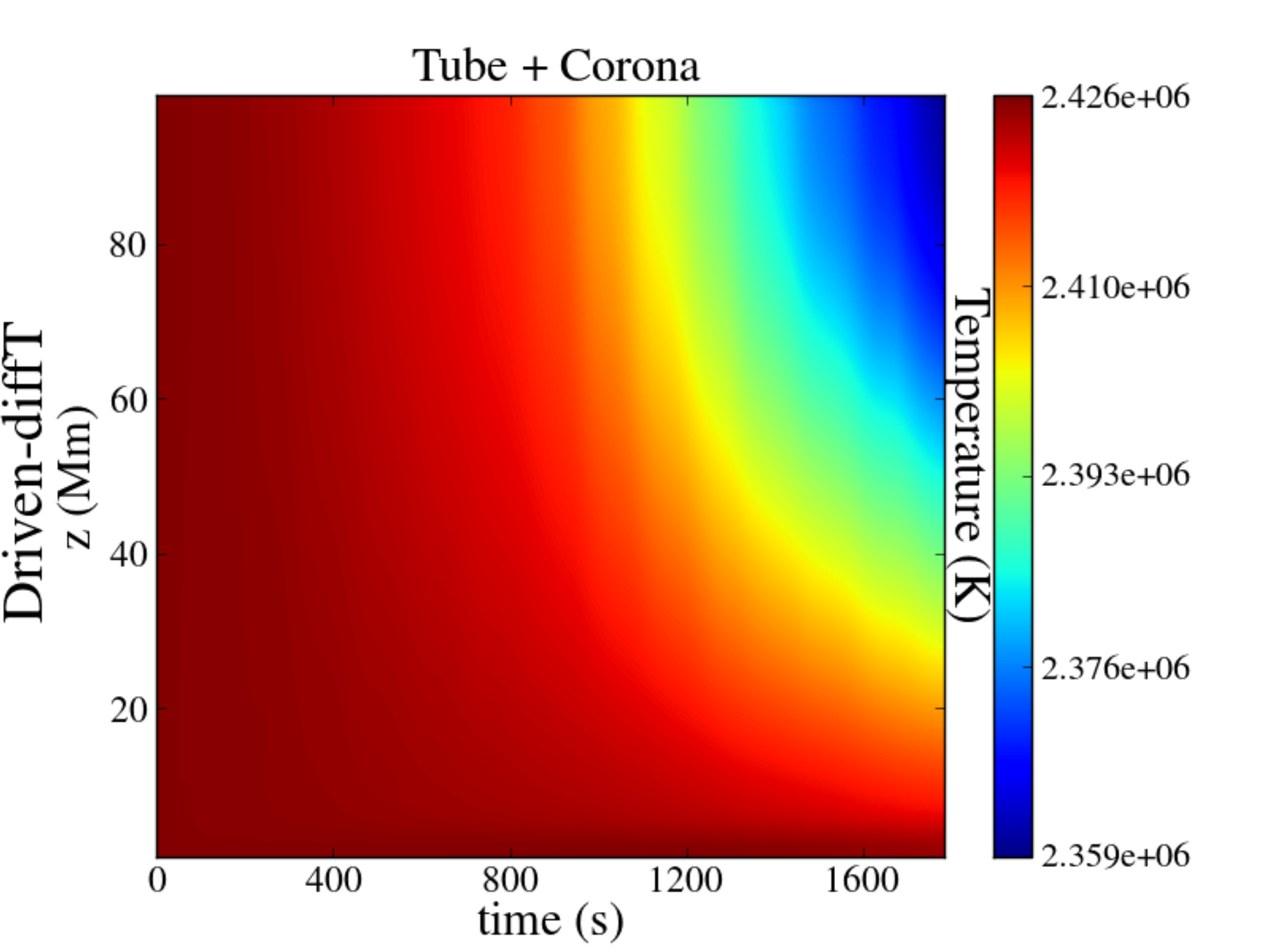}
\includegraphics[trim={1.0cm 0.9cm 0cm 0cm},clip,scale=0.3]{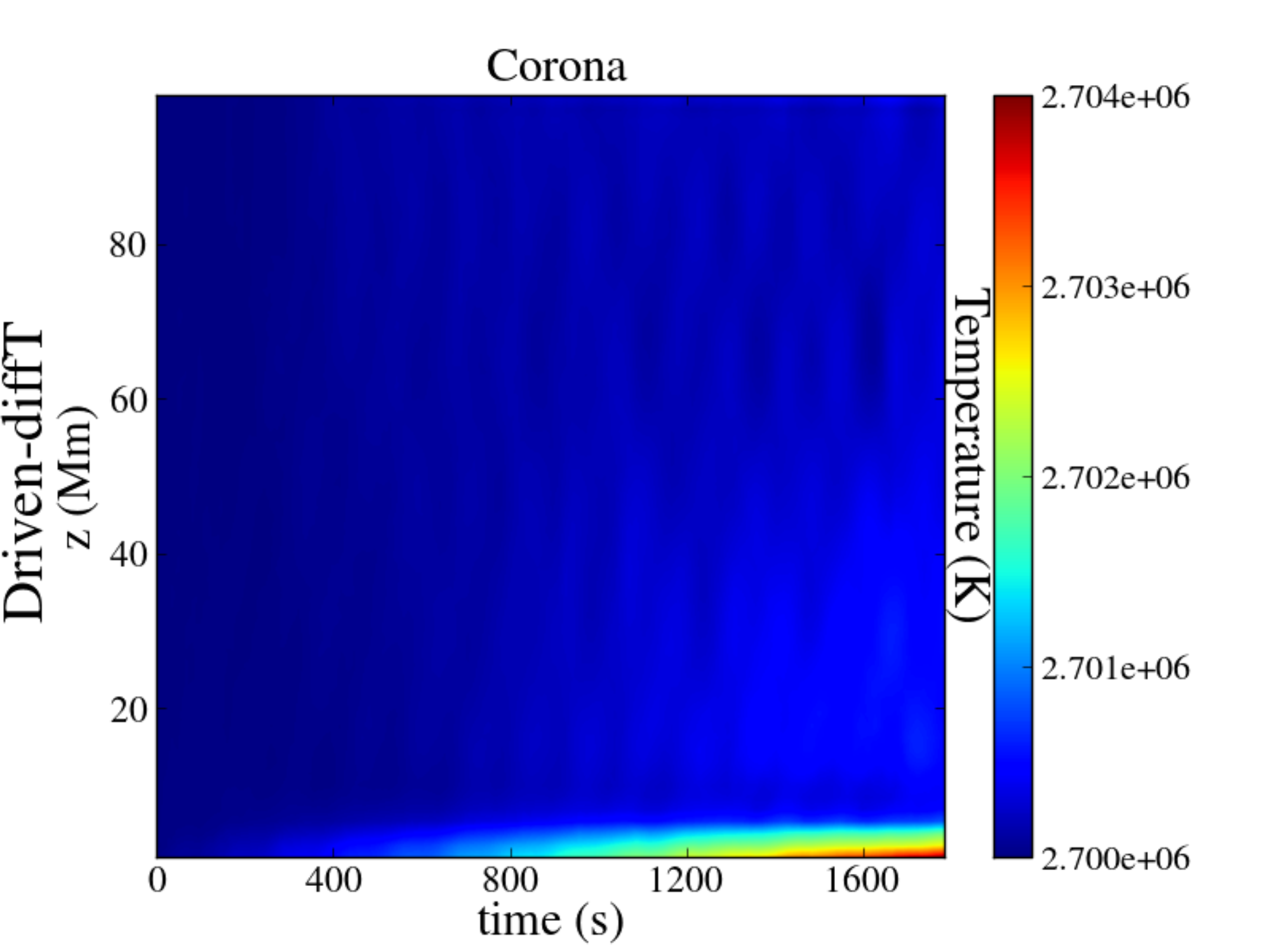}

\includegraphics[trim={1.0cm 0.9cm 0cm 0cm},clip,scale=0.3]{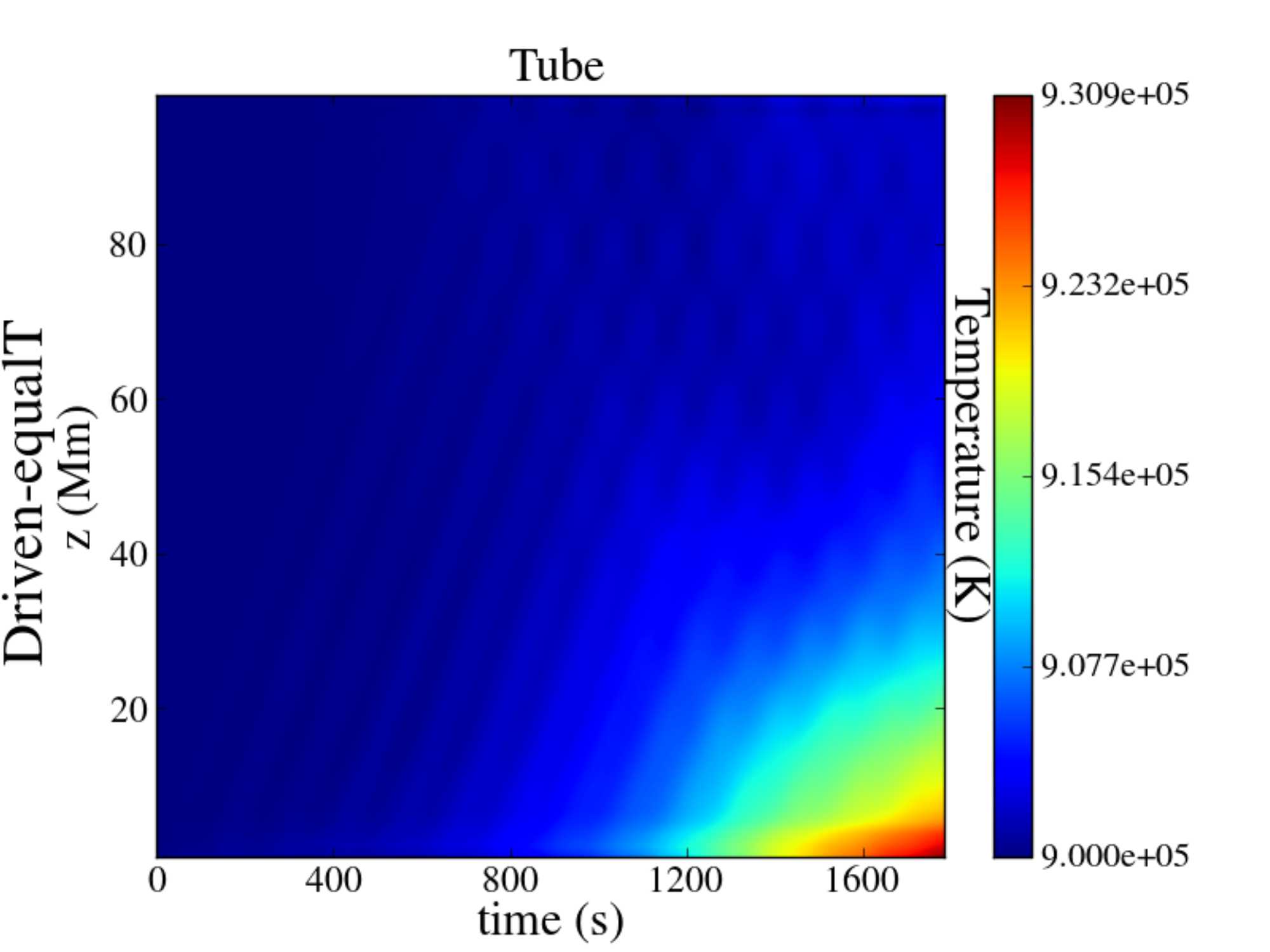}
\includegraphics[trim={1.0cm 0.9cm 0cm 0cm},clip,scale=0.3]{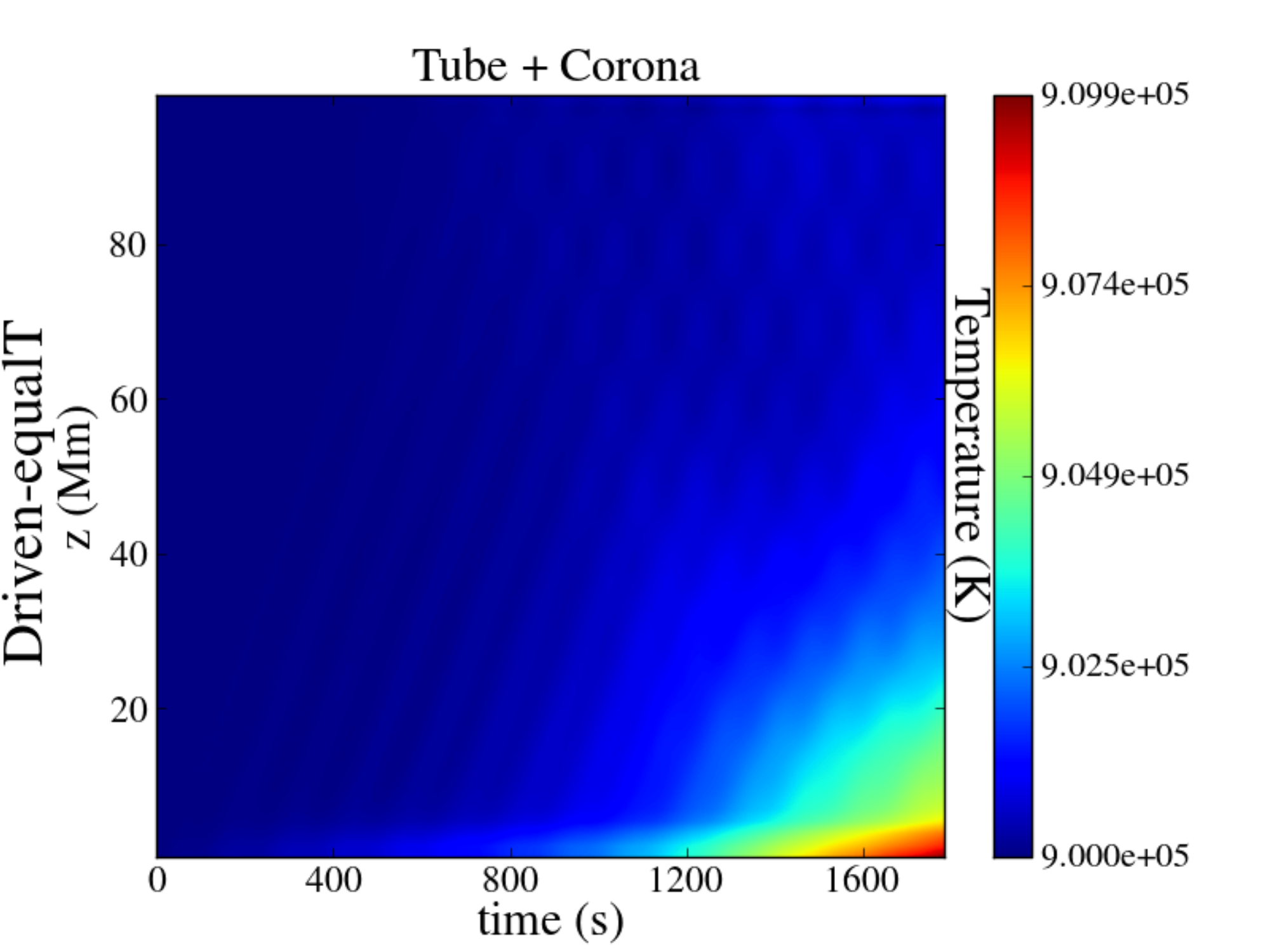}
\includegraphics[trim={1.0cm 0.9cm 0cm 0cm},clip,scale=0.3]{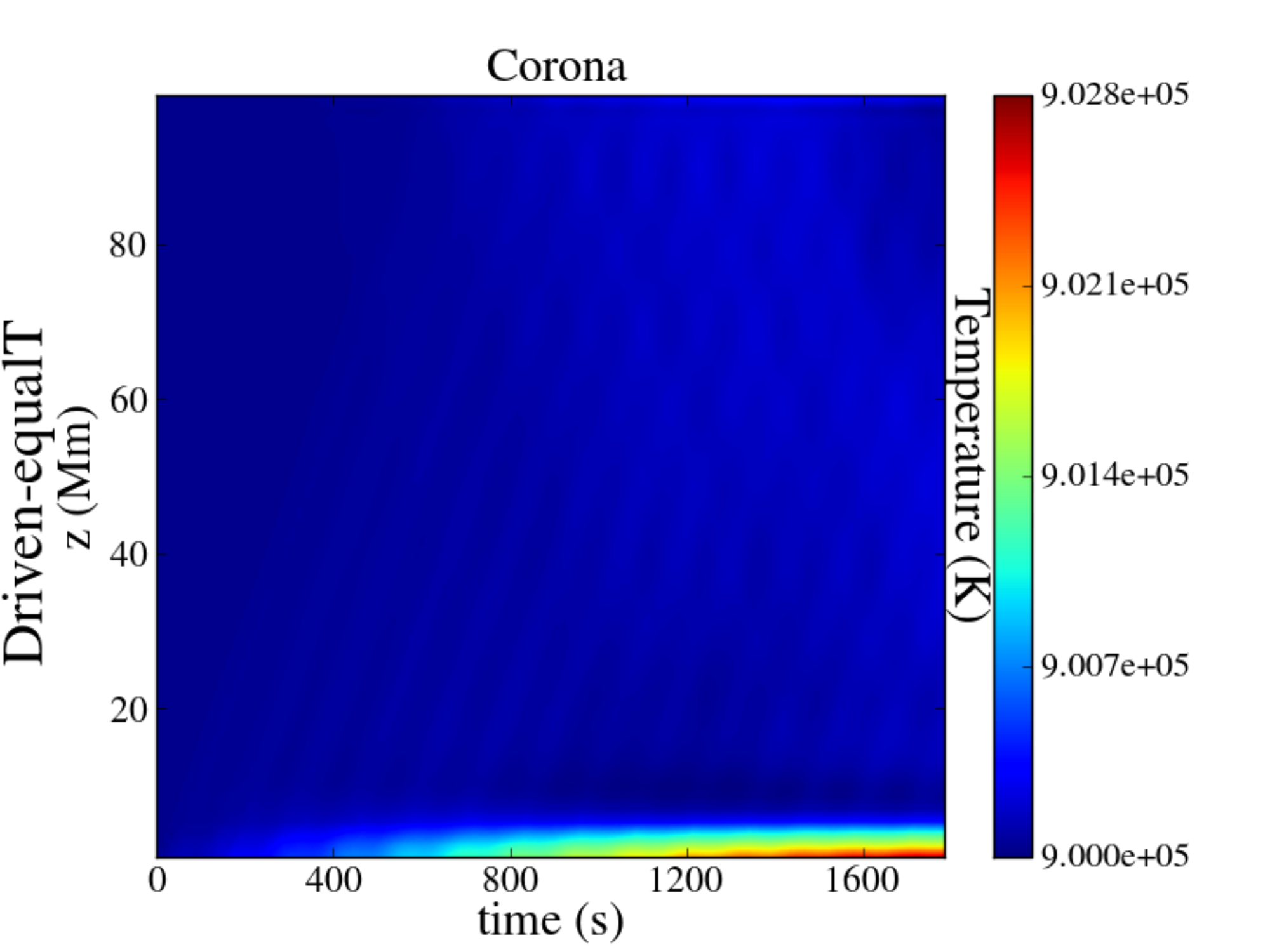}

\includegraphics[trim={1.0cm 0cm 0cm 0cm},clip,scale=0.3]{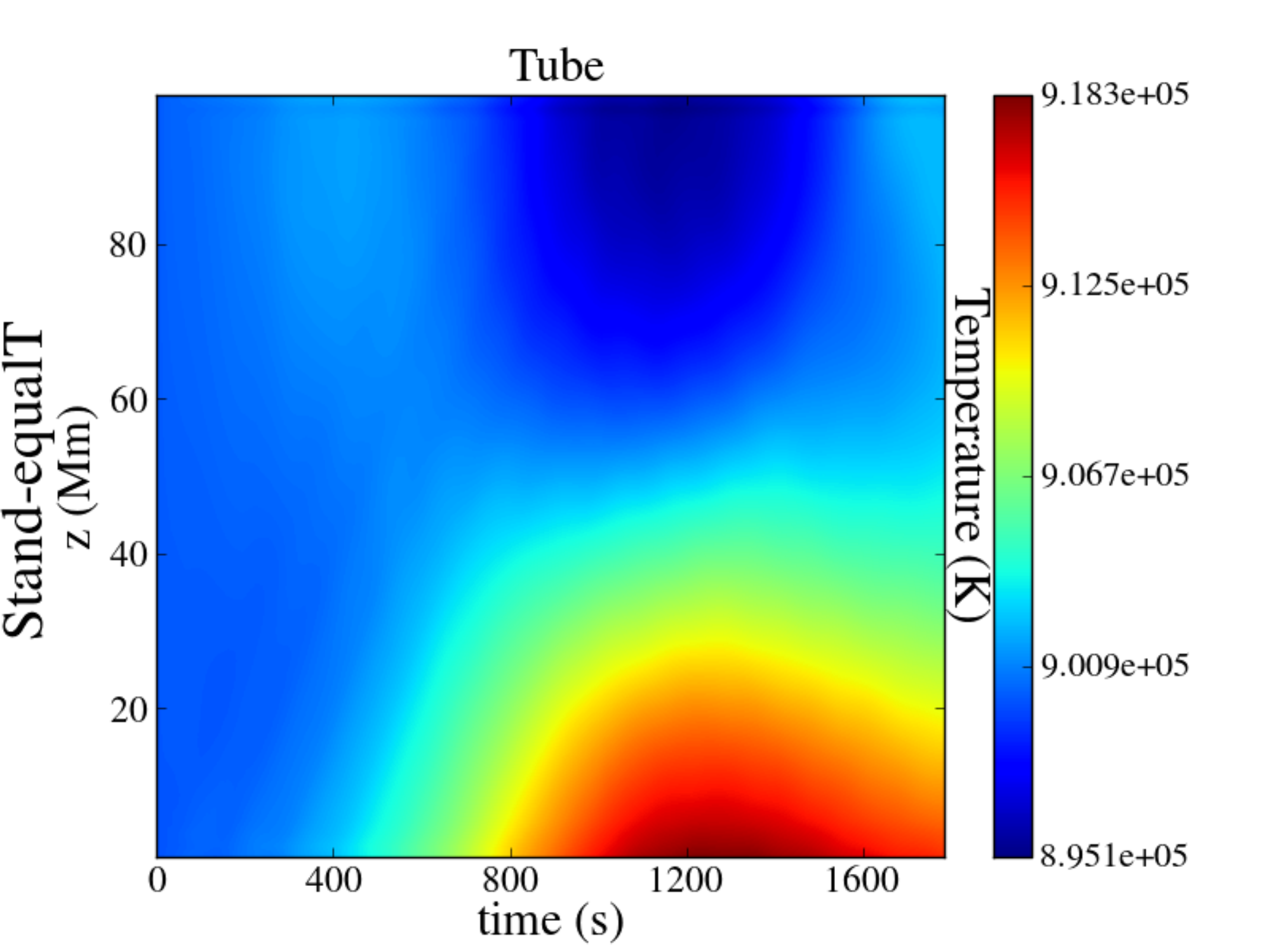}
\includegraphics[trim={1.0cm 0cm 0cm 0cm},clip,scale=0.3]{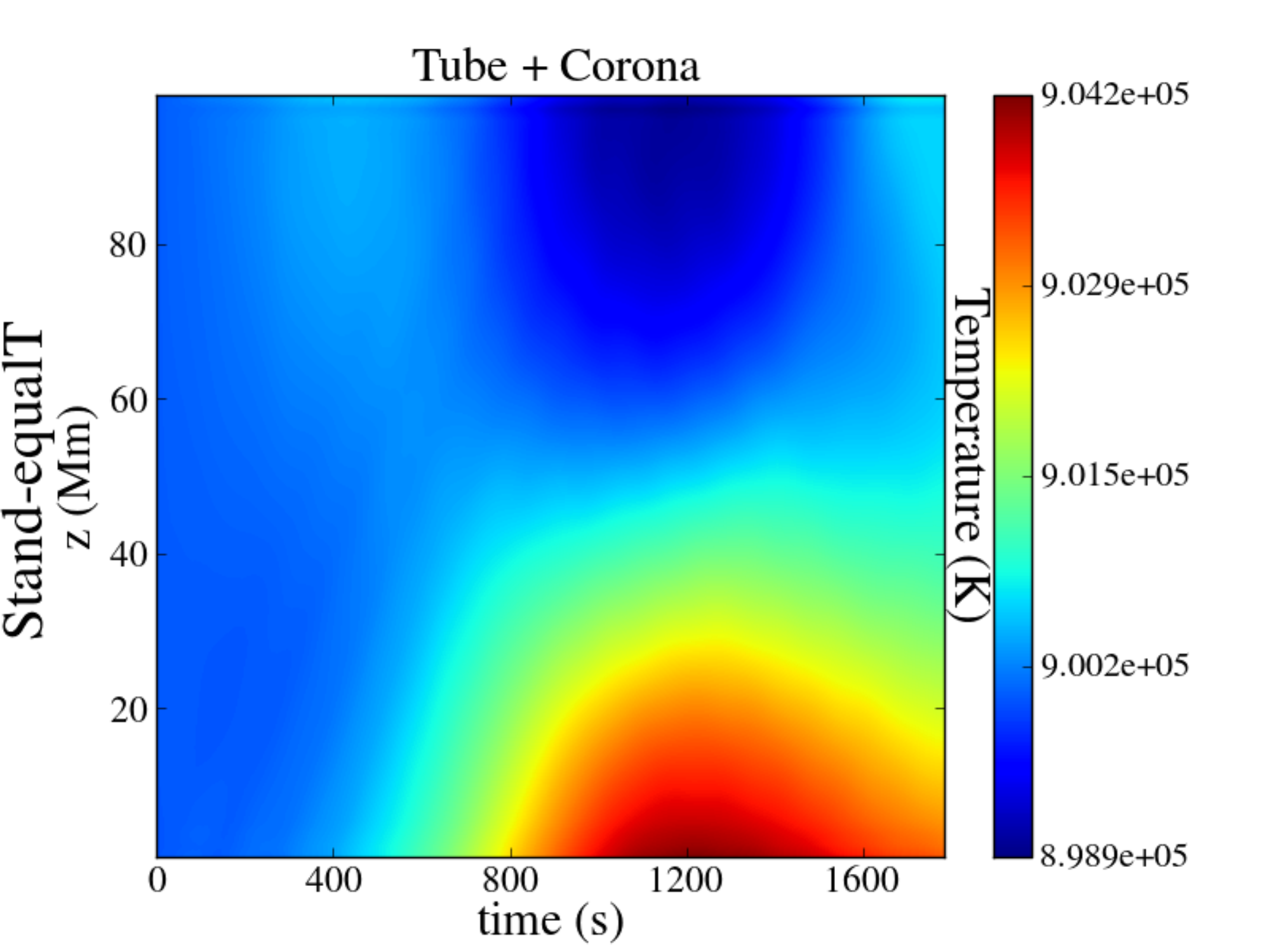}
\includegraphics[trim={1.0cm 0cm 0cm 0cm},clip,scale=0.3]{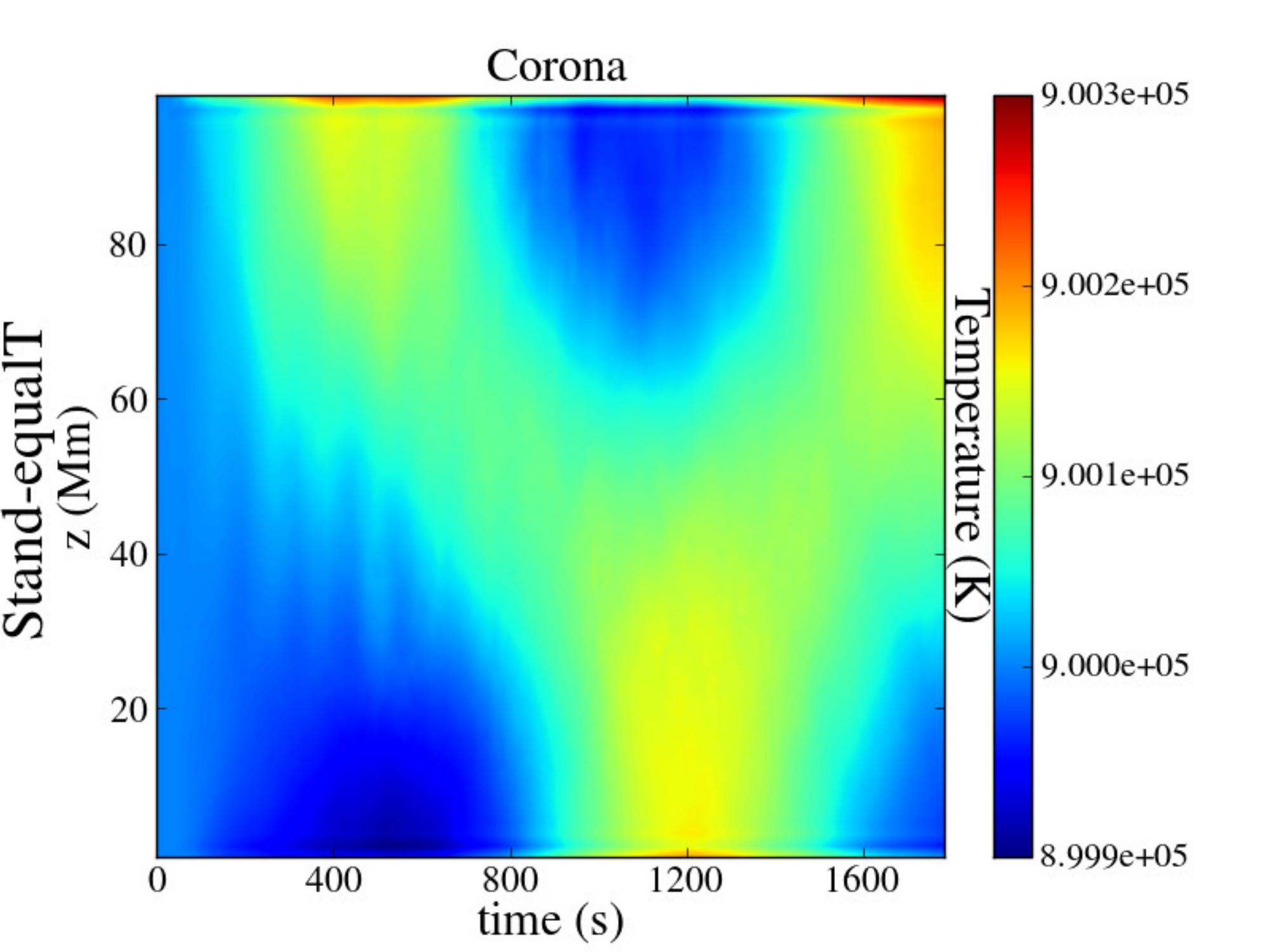}
\caption{Average temperature along the  z-axis: (left) inside the tube (for $\rho \geq 0.84 \cdot 10^{-12}$ kg/m$^3$), (centre) for the whole greater region (tube+corona) including the loop ($0 \leq z \leq 100 $ Mm, $\vert x \vert \leq 2.33$ Mm and $y \leq 2.33$ Mm) and (right) for the surrounding plasma outside the loop. From top to bottom: (a) the Driven-diffT model, (b ) the Driven-equalT model and (c) the Stand-equalT model. The apex is located at $z=100$ Mm.}\label{fig:Tzt}
\end{figure*}

To further study the differences in the internal energy - temperature connection among our models, we will examine the temperature profiles along the $z$-axis, over time, in Fig. ~\ref{fig:Tzt}. For the model of the driven standing wave inside the tube in temperature equilibrium, we observe a gradual increase of the average temperature over time ($\sim 3.4 \% $), the closer we get to the footpoint. The temperature towards the apex increases as well, but by a relatively smaller amount. A similar behaviour is obtained by the temperature profiles of both the greater region (tube + corona) and for the surrounding corona. Regarding this model, the lower changes in the average temperature for the `tube + corona' area and for the surrounding corona are attributed to the larger area studied, while the heating is located only inside and at the boundary of the loop. This is also in agreement with our findings in Fig. \ref{fig:ietemps} for this model.

Studying the corresponding temperature profiles for the standing oscillating loop, we observe again the highest average temperatures towards the footpoints. The new phenomenon that we did not encounter in the previous case, is a long period oscillation of the average temperature. This oscillation, mostly prominent at the apex, is due to the longitudinal slow mode triggered by the large initial velocity perturbation \citep{terradas2011A&Aslowmode, magyar2016damping}. In the case of the driven waves, this mode is not observed, the reason being probably the gradual energy input from the footpoint. Despite the presence of this mode, however, the temperature profile for the tube oscillating as a standing wave still indicates heating along a large part of the loop, with the highest temperatures seen towards the footpoint. As before, the same temperature evolution is observed in the greater region, whereas a similar trend is observed in the surrounding corona, although affected by the effects of the long period slow mode.

\begin{figure*}
\centering
\resizebox{\hsize}{!}{\includegraphics[trim={0cm 0cm 1cm 0cm},clip,scale=0.3]{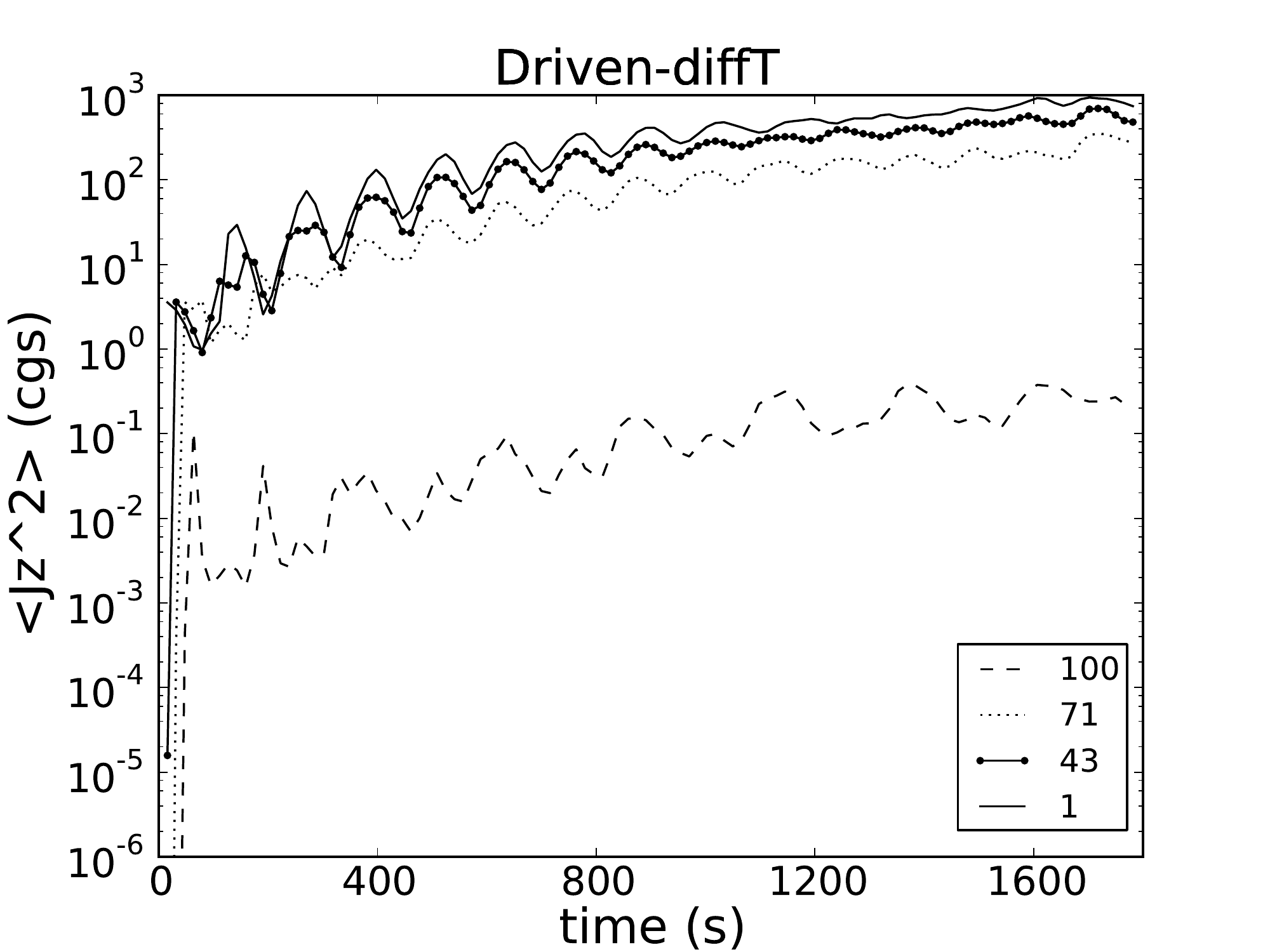}
\includegraphics[trim={0cm 0cm 1cm 0cm},clip,scale=0.3]{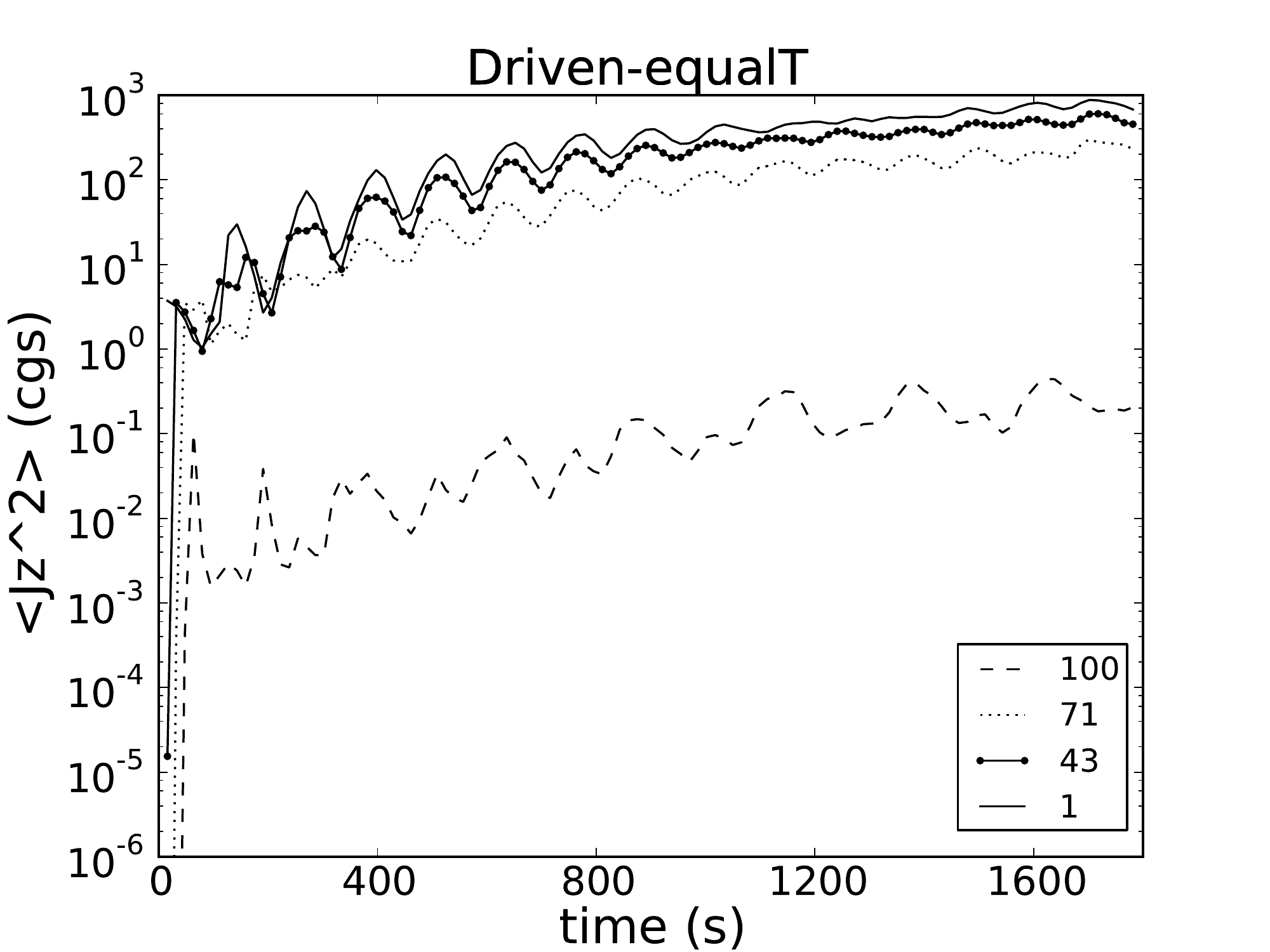}
\includegraphics[trim={0cm 0cm 1cm 0cm},clip,scale=0.3]{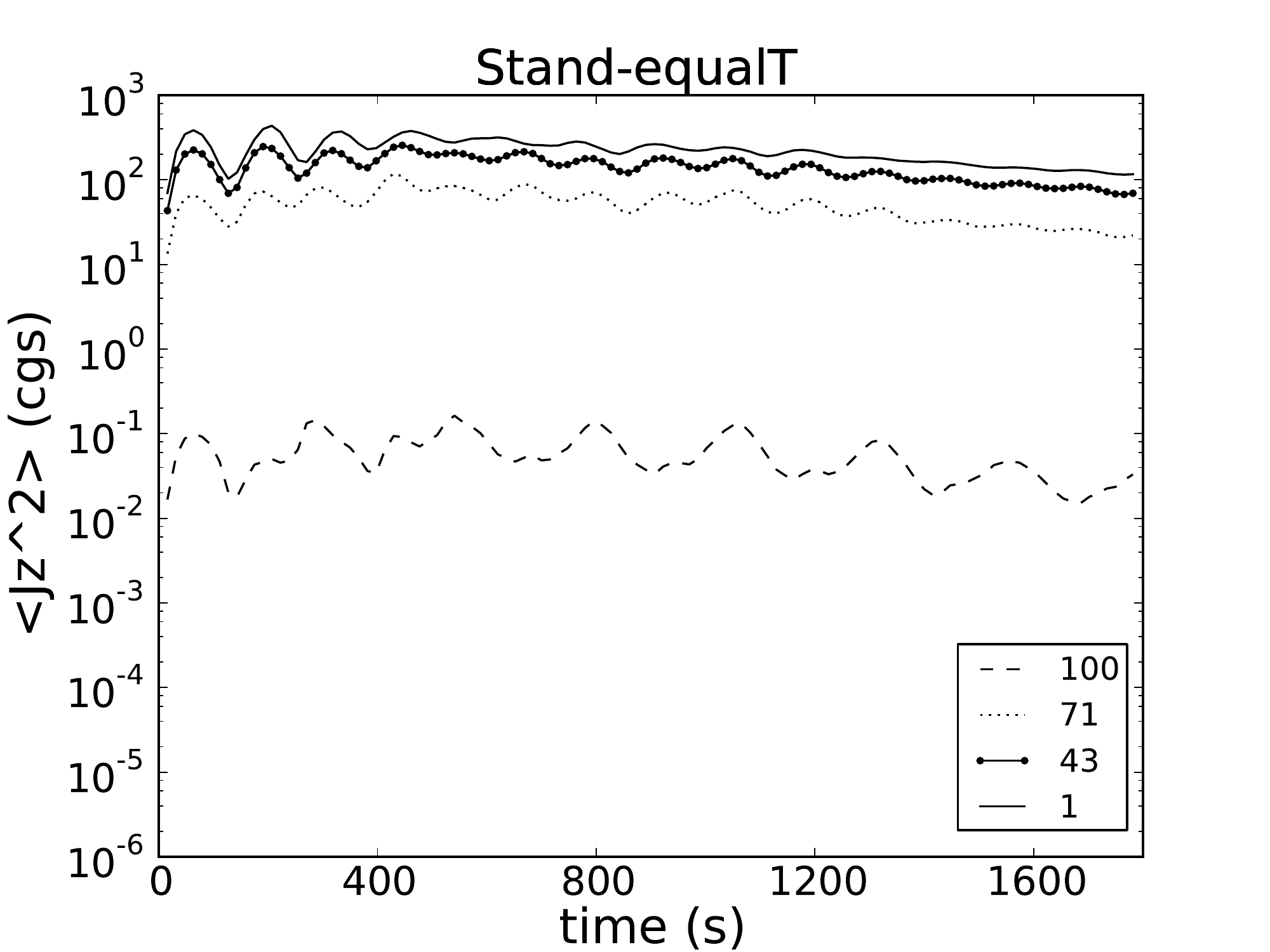}}

\resizebox{\hsize}{!}{\includegraphics[trim={0cm 0cm 1cm 0cm},clip,scale=0.3]{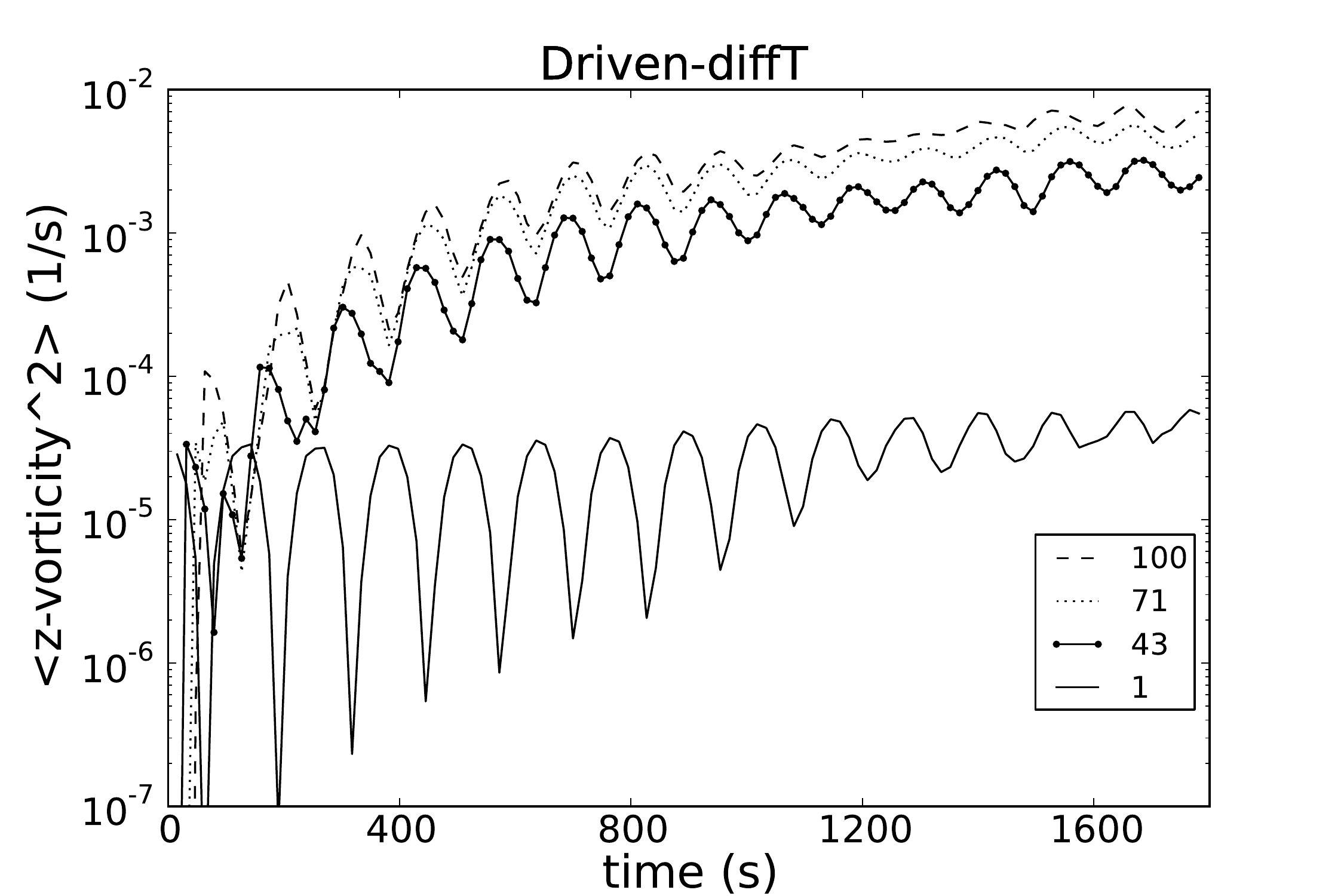}
\includegraphics[trim={0cm 0cm 1cm 0cm},clip,scale=0.3]{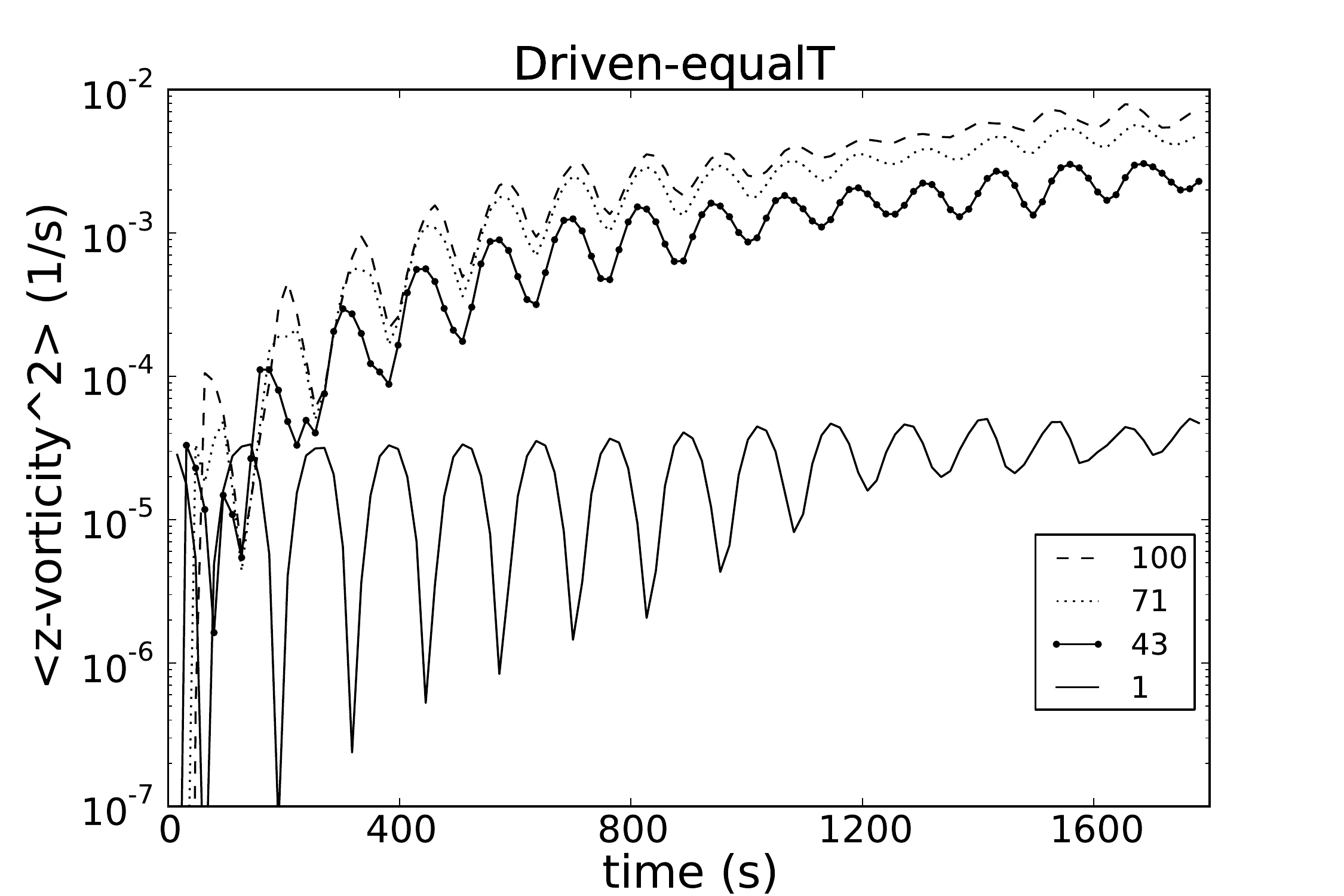}
\includegraphics[trim={0cm 0cm 1cm 0cm},clip,scale=0.3]{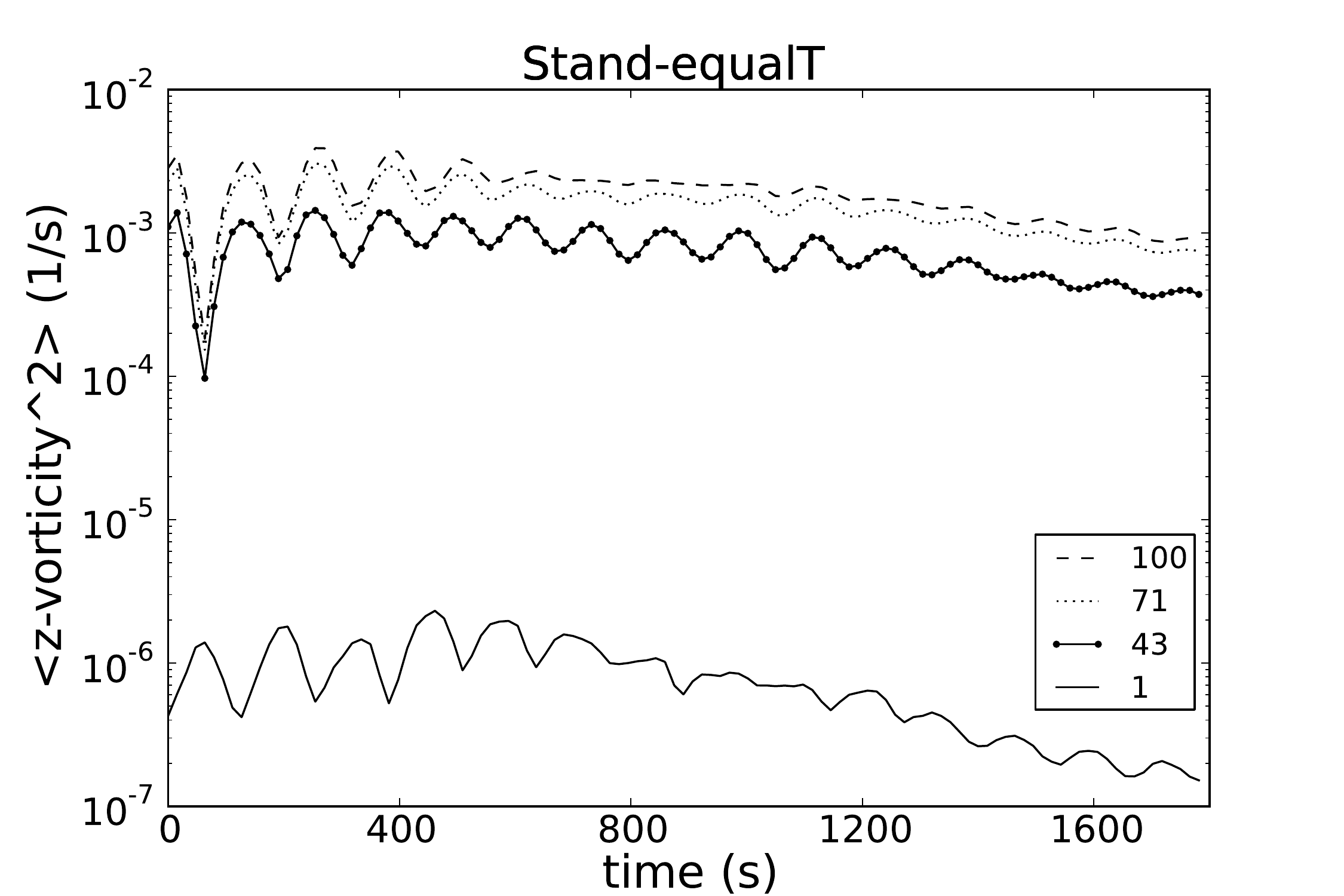}}
\caption{Top and bottom panels correspond to time profiles for the average square $z$-current density $J_z^2$ and the average square $z$-vorticity $\omega_z^2$, respectively, at different heights. The apex is located at $z=100$ Mm and the volume averaging takes place only inside each tube ($ \rho > 0.335 \, \rho_i$).  From left to right: (a) the Driven-diffT model, (b) the Driven-equalT model and (c) the Stand-equalT model.}\label{fig:curvort}
\end{figure*}

Before addressing the Driven-diffT model, we will try to recognize the different dissipation mechanisms involved by plotting, in Fig.~\ref{fig:curvort}, the average square $z$-vorticity ($\omega_z^2$) and the average square $z$-current density ($J_z^2$) for different heights inside our three different loops. These plots are focused on the area inside the tubes, defined by density $ \rho > 0.335 \, \rho_i$. It is obvious from all three models that the square $z$-vorticity gains its highest values at the apex, which is the position of the velocity antinode and the part of the loop where the Kelvin-Helmholtz generated turbulence is the strongest. The higher velocities observed there, due to the larger amplitude of the oscillation, lead to greater values of the squared $z$-vorticity. This can also be verified by the periodicity of $\omega_z^2$. Its period is half that of the wave, which gives a period for the $\omega_z$ vector equal to that of the wave. Also, for the Stand-equalT model, we observe a drop of the vorticity oscillation amplitude, as well as of its mean value. This is mainly due to damping of the kink wave. The square $z$-vorticity also seems to be connected to the induced turbulence in our tubes. As the tubes start oscillating, the minimum values of $\omega_z^2$ grow and gain non-zero values through the creation of smaller scale structures (like TWIKH rolls). This, `base value' is larger at the apex, where the KHI, and the induced turbulence, are at their strongest. Therefore, $\omega_z^2$ can be used as a measure of turbulence.

The $z$-current density, however, is increasing towards the footpoint, where it becomes three orders of magnitude larger than at the apex, for all three models. This consistency is caused by the geometry of the resulting oscillations. As we have already mentioned, after the superposition of the counter-propagating waves from each footpoint, we have the manifestation of a footpoint driven standing wave which resembles the fundamental standing kink mode that the tube. By writing the magnetic field and the current density in cylindrical coordinates, we see that the main contribution on $J_z$ is from the radial variation of azimuthal component of the magnetic field $\partial B_{\phi} / \partial r$, which for the case of the fundamental kink mode has a cosine dependence along the $z$ axis, following $B_{\phi}$. Thus, the z-current density will get its highest values near the footpoints of such oscillating loops \citep{tvd2007resist}. As we mentioned in the introduction, in that paper, \citeauthor{tvd2007resist} have proven that for line-tied loops the viscous and resistive heating mechanisms can be observationally distinguished by the site of the heating. Ohmic dissipation, due to resistivity, is more prominent at the footpoints of oscillating loops, while viscous dissipation is stronger towards the apex. Combining the above with the value of the resistive time scale for our models, we conclude that in both equalT-models, the stronger rise of temperature at the footpoint is an indication of Ohmic heating due to numerical resistivity. The lower increase of the average temperature at the apex could be potentially caused by viscous dissipation. However, the temperature gradient between the apex and the footpoint, for these two models, suggests that resistive heating is the dominant heating mechanism in our models.

\begin{figure}
\resizebox{\hsize}{!}{\includegraphics[trim={0.5cm 0cm 1.4cm 0cm},clip,scale=0.23]{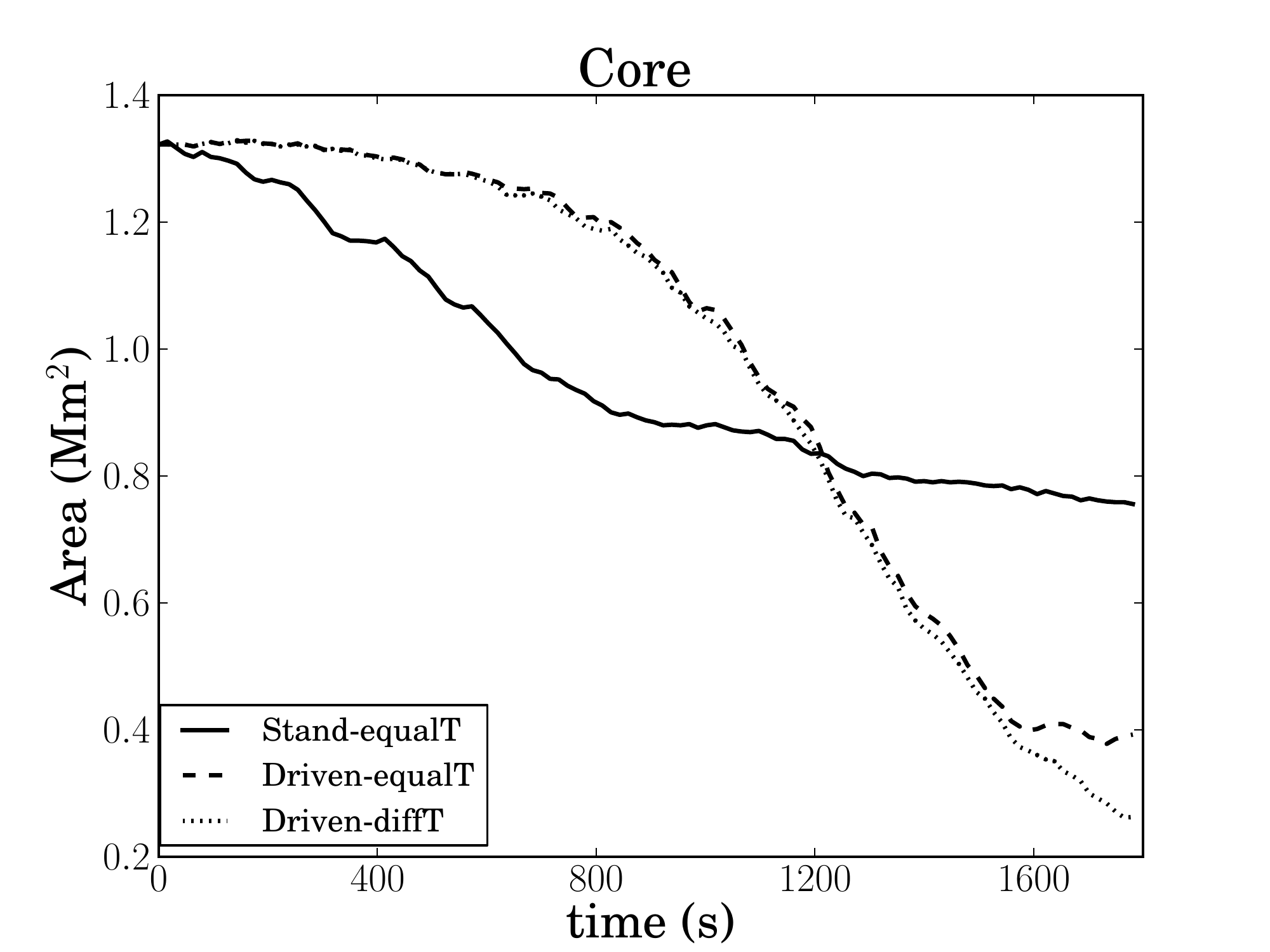}
\includegraphics[trim={0.5cm 0cm 1.4cm 0cm},clip,scale=0.23]{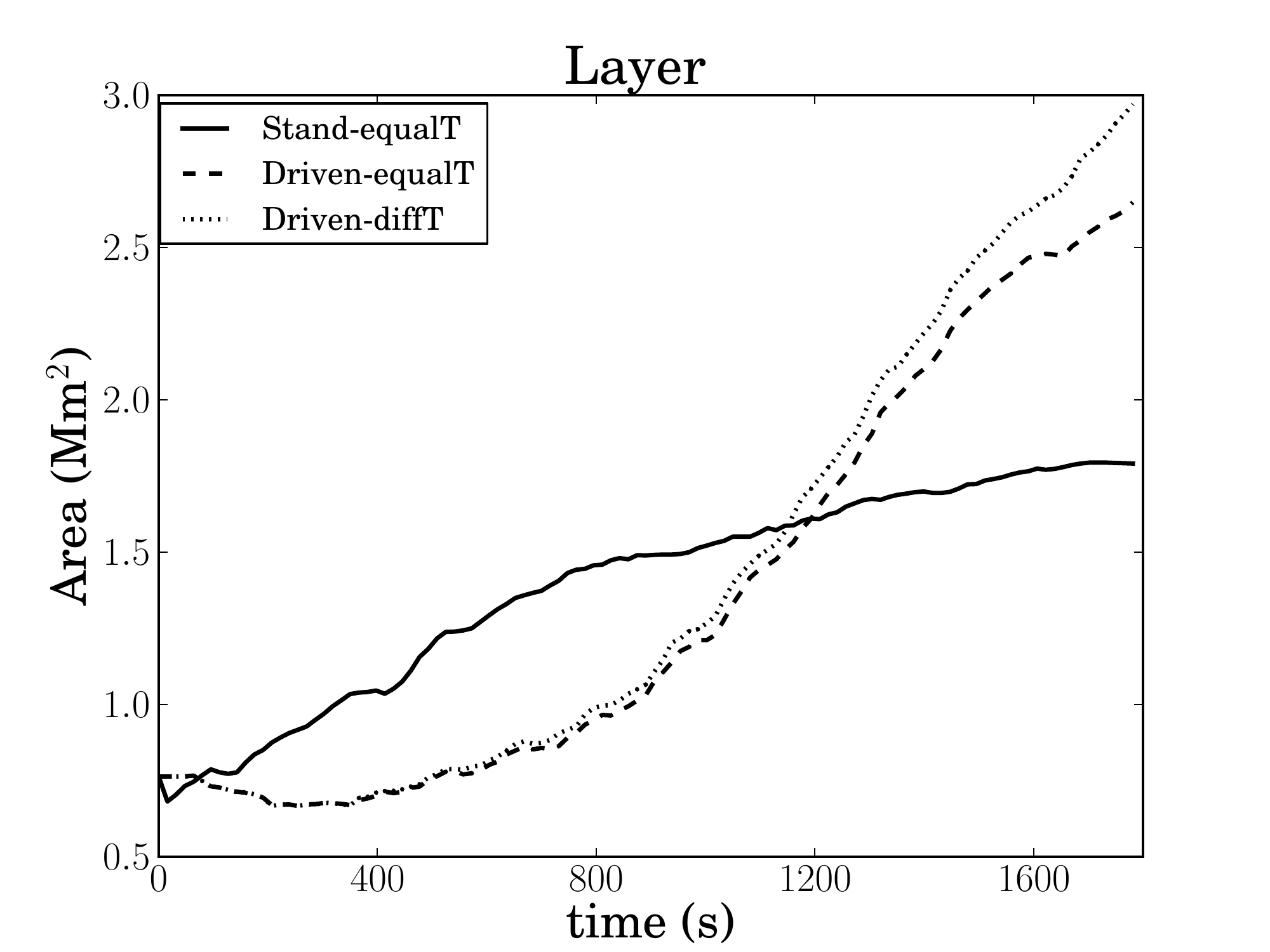}}
\resizebox{\hsize}{!}{\includegraphics[trim={0.5cm 0cm 1.4cm 0cm},clip,scale=0.23]{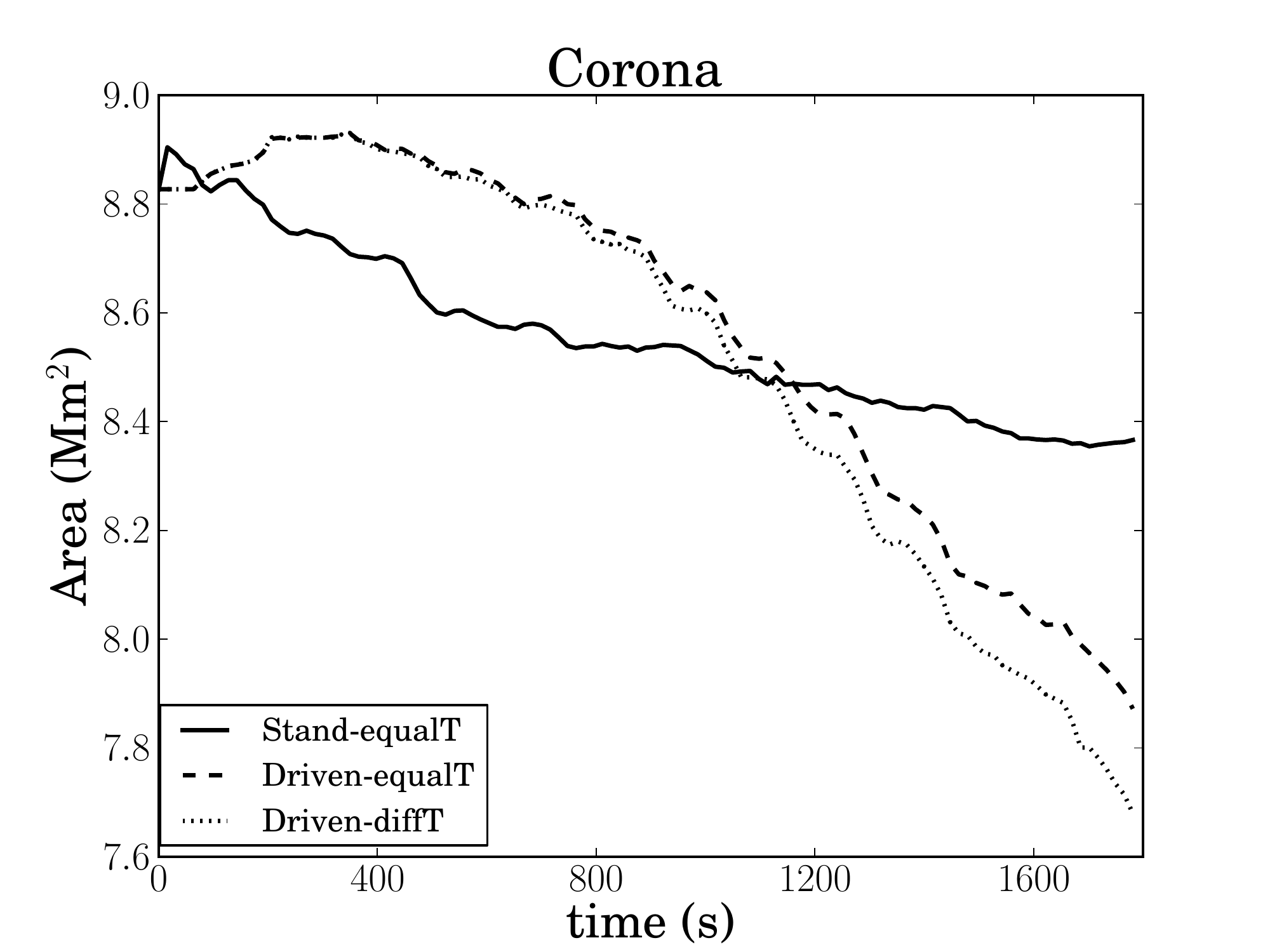}
\includegraphics[trim={0.5cm 0cm 1.4cm 0cm},clip,scale=0.23]{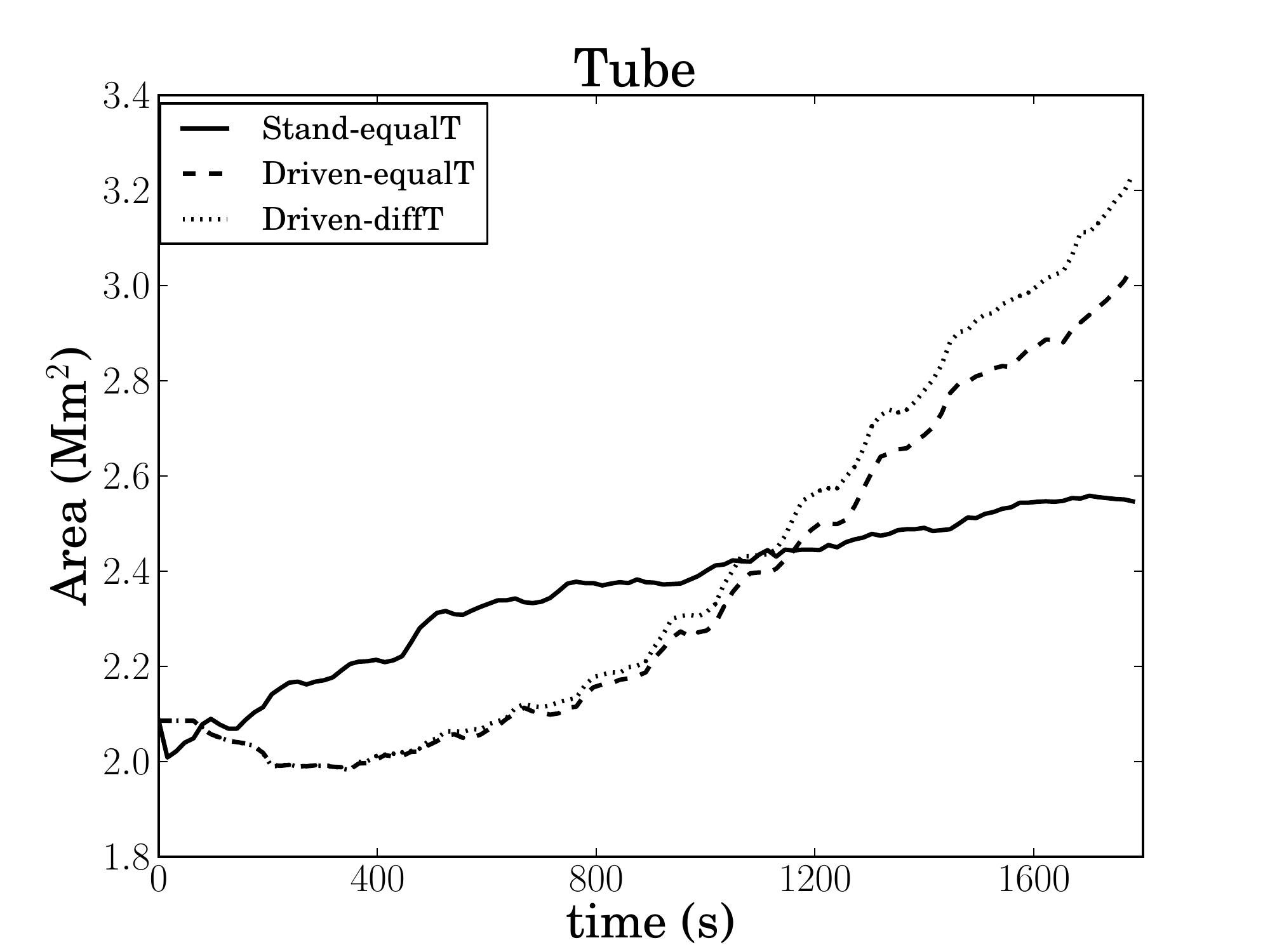}}
\caption{Time evolution of the core ($\rho \geqslant 0.976 \, \rho_i$), layer, corona ($\rho \leqslant 0.335 \, \rho_i$) and whole tube ($\rho > 0.335 \, \rho_i$) surface areas (in $Mm^2$) for our three loops, at the apex. The continuous line represents the Stand-equalT model, the dashed line represents the Driven-equalT model and the dotted line the Driven-diffT model.}\label{fig:areas}
\end{figure}

For the model of the driven wave inside the tube with a temperature gradient, from Fig. ~\ref{fig:Tzt}, we observe a gradual increase of the average temperature over time, the closer we get to the apex. This phenomenon seems to contradict the results we have got so far about the wave heating mechanisms, since this model is dynamically the same as the model of propagating waves inside the tube in temperature equilibrium. It is however consistent with the findings of \citet{magyar2016damping} for the standing kink, where it was observed that the internal energy rise at the layer was not enough to explain the rise of the temperature at the layer. Initially, as we see in Fig.~\ref{fig:areas}, for this model, the hotter boundary layer ($0.335\, \rho_i < \rho < 0.976 \, \rho_i $) shrinks over time. The overall volume of the tube is reduced and an initial drop of the volume average temperature inside the tube is caused, since the relative contribution from the colder inner parts of the loop increases. As the simulation runs, however, the development of the turbulence and the manifestation of the TWIKH rolls lead to extensive mixing with the surrounding area, expanding the turbulent layer both inwards and outwards. The shrinking of the colder core region and the expansion of the hot turbulent layer is what causes the tube to become hotter over time. It is no coincidence that the greatest rise of temperature takes place at the apex, where the average vorticity (or in our case the average square vorticity) takes its highest values, as seen in Fig. ~\ref{fig:curvort}. The apex is the location of the velocity antinode, where the oscillation amplitude and the induced turbulence are the strongest. Therefore, the mixing is also more extensive there, increasing the temperature of the loop even further. The footpoint ($z=0$) does not show any rise in temperature. On the contrary, the shrinking of the hotter layer, and thus the increase of the colder core contribution, drops the temperature to a point where Ohmic heating, which is also present in this model, is not adequate to sustain the initial average temperature. 

Studying the greater area ($0 \leq z \leq 100 $ Mm, $\vert x \vert \leq 2.33$ Mm and $y \leq 2.33$ Mm) for the Driven-diffT model in Fig.~\ref{fig:Tzt}, we observe a temperature drop close to the apex. This behaviour, which is also connected to our findings in Fig. \ref{fig:ietemps}, is caused by the extensive mixing between the loop and the hotter (in this model) environment. The temperature profile has the opposite evolution than before, decreasing near the apex and increasing near the footpoint. This is a confirmation of the assumption about the mixing. Through the loop expansion, both the colder core as well as the warm layer contribute more to the average temperature, than they initially did. Both the layer and the core are colder than the surrounding plasma. This causes the volume average temperature to drop. However, this does not mean that the environment cools down. From the temperature spatial profiles in Fig. \ref{fig:TRHO}, we can see a rise in temperature of the surrounding plasma, as we approach the turbulent layer. Plotting the average temperature along the $z$-axis over time (Fig. ~\ref{fig:Tzt}), for the surrounding area, this slight heating of the corona becomes obvious. It is worth noting that even in this case, the footpoint reaches slightly higher temperatures than the apex, which again can be explained through the effects of resistive heating at the tube-corona interface, at the edges of the boundary layer. 

The Driven-equalT model also exhibits the same evolution regarding the different areas studied, but unlike the Driven-diffT model, the changes of the tube cross-section do not affect the average temperature of each region. Instead, the wave dissipation mechanisms, in particular resistive heating, are the ones responsible for the temperature fluctuations. The same is valid also for the tube oscillating as a standing wave. Finally, we  stress that the results of the mixing on the apparent heating (cooling) of the tube (tube + environment) depend on the temperature gradient between the environment and the loop; they could significantly change should we consider different initial temperature gradients.  

\section{Conclusions}

We were interested in studying the heating produced by transverse waves in coronal loops. We performed numerical simulations for a 3D, density enhanced straight tube in ideal MHD. The effective value of numerical resistivity present in our model is many orders of magnitude larger than the expected values in the solar corona. We studied two models of driven standing waves, for a continuous, monoperiodic, footpoint driver. One model had a uniform temperature throughout the domain, while the other had a temperature difference between the loop and the environment. The velocity amplitude of our driver was of the order of $2$ km/s, while the period of the driver was equal to the analytically predicted value for the standing fundamental kink oscillations of a uniform flux tube for our given densities \citep{edwin1983wave}. By choosing that corresponding frequency for our driver, and considering symmetry at the apex, the initially generated propagating waves superpose and form the fundamental standing mode of a kink oscillation. As predicted by the theory of driven mechanical oscillations, the original increase of our oscillation amplitude and of the corresponding $v_x$ velocity, due to the continuous input of energy, gave way to a saturation point in both models. The noticeable difference in the magnetic energy density profiles between the two models did not seem to affect the dynamics of the two systems. Notably, the internal energy density variation relative to the initial state, as well as the kinetic energy density were similar in both models. Additionally, considering a uniform temperature model of a tube oscillating as a standing wave, with an initially sinusoidal perturbation in the velocity, we reproduced a damping profile similar to those in \citet{magyar2016damping}. 

For all three of our models, we reported the creation of turbulence at the edges of our loop and the development of Transverse Waves Induced Kelvin-Helmholtz (TWIKH) rolls. These rolls resulted in extensive mixing of plasma between the inner loop and the surrounding corona, as shown in the profiles of temperature ($T$) and density ($\rho$) at the apex. By considering an initial temperature equilibrium between the loop and the environment, we see that the KHI produces both a temperature increase and a decrease in the turbulent layer. These temperature fluctuations take place at the location of the TWIKH rolls and, as reported by \citet{antolin2017arXiv170200775A}, they are not caused by the transfer of energy between the different TWIKH rolls. Instead they are connected to the pressure and density fluctuations caused by the turbulence, thus being adiabatic in nature. Similar temperature fluctuations can be observed at the footpoints of our uniform-temperature models as well, likely caused by the large scale dynamics due to the waves examined. 

In the two models with the uniform initial temperature, the increase of the volume averaged energy density of the tubes was the same in percentage to the corresponding rise in temperature. This proved that there is indeed a wave dissipation mechanism that causes conversion of magnetic and kinetic energy to thermal energy. Studying the temperature profiles along the loop axis and over time, we observed a site of heating near the loop footpoints that is present both in the case of the driven standing wave and the impulsively excited standing wave. The profiles of the square volume averaged $z$-current density near the footpoints (three orders of magnitude higher at the footpoint than at the apex), indicated a strong contribution of Ohmic heating, due to numerical resistivity. A careful study of the temperature profiles for the case of the driven oscillation indicated a slight increase of temperature near the apex, were the turbulence is the strongest, and the square $z$-vorticity gets its highest values for all of our models. These higher values near the apex, are due to both the higher velocities encountered there and the creation of more prominent smaller scale structures, such as TWIKH rolls. Further studies must be done in order to estimate the contribution from viscous heating there, as well as the effects of actual, physical, resistivity. However, the greater increase of the average temperature that is observed near the footpoints, is caused by the currents generated at the turbulent layer. This suggests that resistive dissipation is the main mechanism for heating \citep{tvd2007resist}.

For the model without the driver, we end up with the same preference of resistive over viscous heating, with the highest temperatures and z-current densities observed near the footpoint. The observed long period oscillation of the temperature in this case, is due to the longitudinal slow mode initiated, triggered by initial the perturbation \citep{magyar2016damping}. In the case of the driven standing waves, this mode was not observed in our models, probably because of the gradual energy input from the footpoints. The overall lower available energy from the standing oscillation, alongside the effects of the observed slow mode, eventually led to a lower increase of the loop volume averaged temperature. In comparison, despite their slower initial heating, the driven waves produced higher temperatures after the oscillations entered their assumed saturation point.

Finally, testing the effects of temperature variation between the flux tube and the environment for the driven oscillations, we observed that the perturbation in the internal energy and the evolution of the temperature follow different profiles over time. We reached the conclusion, that extensive mixing between plasmas of different temperatures can potentially hide the effects of the wave heating mechanisms. This apparent heating (or cooling, not considered here) is generally determined by the initial temperature difference between the flux tube and the environment, meaning that varying results should be expected for different gradients. This has to be taken into account when dealing with observations, since a higher calculated temperature would not necessarily mean actual heating of the whole loop-atmosphere system. In our model of propagating waves of a cold tube in a hotter environment, the rise of temperature was the highest at the apex, where the $z$-vorticity also took its highest values. The resulting temperatures were far greater than those produced by the wave heating in the model of temperature equilibrium, in agreement with \citet{magyar2016damping}. 
\\

\begin{small}
\textit{Acknowledgements.} We would like to thank the referee, whose comments led to a great improvement of the manuscript. We also thank the editor, for his suggestions. K.K. was funded by GOA-2015-014 (KU Leuven). T.V.D was supported by the IAP P7/08 CHARM (Belspo) and the GOA-2015-014 (KU Leuven). P.A. acknowledges funding from the UK Science and Technology Facilities Council and the European Union Horizon 2020 research and innovation programme (grant agreement No. 647214). The results were inspired by discussions at the ISSI-Bern and at ISSI-Beijing meetings. 
\end{small}

\bibliographystyle{aa}
\bibliography{tubeheatingaa}

\end{document}